\documentclass[%
reprint,
amsmath,amssymb,
prl,superscriptaddress
]{revtex4-2}
\usepackage[utf8]{inputenc}
\usepackage{graphicx}
\usepackage{dcolumn}
\usepackage{bm}
\usepackage{mathtools}
\usepackage{color}

\DeclareMathOperator{\Tr}{Tr}

\DeclareMathOperator{\conv2D}{\ast\ast}
\DeclarePairedDelimiterX{\abs}[1]{\vert}{\vert}{#1}
\DeclarePairedDelimiterX{\norm}[1]{\lVert}{\rVert}{#1}
\DeclarePairedDelimiterX{\expval}[1]{\langle}{\rangle}{#1}
\DeclarePairedDelimiterX{\ket}[1]{\vert}{\rangle}{#1}
\DeclarePairedDelimiterX{\bra}[1]{\langle}{\vert}{#1}
\DeclarePairedDelimiterX{\innerproduct}[2]{\langle}{\rangle}{#1\delimsize\vert\mathopen{}#2}
\DeclarePairedDelimiterX{\outerproduct}[2]{\vert}{\vert}{#1\delimsize\rangle\!\delimsize\langle\mathopen{}#2}
\DeclarePairedDelimiterX{\mel}[3]{\langle}{\rangle}%
{#1\delimsize\vert\mathopen{}#2\delimsize\vert\mathopen{}#3}

\begin{document}

\title{Identifying Objects at the Quantum Limit for Super-Resolution Imaging}

\author{Michael R Grace}
\affiliation{College of Optical Sciences, University of Arizona, Tucson, AZ 85721, USA}
\author{Saikat Guha}
\affiliation{College of Optical Sciences, University of Arizona, Tucson, AZ 85721, USA}

\begin{abstract}
	We consider passive imaging tasks involving discrimination between known candidate objects and investigate the best possible accuracy with which the correct object can be identified. We analytically compute quantum-limited error bounds 
	for hypothesis tests on \emph{any} database of incoherent, quasi-monochromatic objects when the imaging system is dominated by optical diffraction. We further show that object-independent linear-optical spatial processing of the collected light exactly achieves these ultimate error rates, exhibiting superior scaling than spatially-resolved direct imaging as the scene becomes more severely diffraction-limited. 
	We apply our results to example imaging scenarios and find conditions under which super-resolution object discrimination can be physically realized.
\end{abstract}

\maketitle

\emph{Introduction}---Object discrimination is at the heart of decision making in medical diagnostics, extrasolar astronomy, and autonomous sensing. For incoherent imaging with large standoff distances, small objects, and/or aperture-limited imaging systems, the physical principle of diffraction
impedes accurate discrimination between spatially distinct objects. A classic heuristic criterion, attributed to Rayleigh, holds that two objects cannot be discriminated when their distinguishing features exhibit length scales smaller than the width of the system point spread function (PSF)~\cite{Rayleigh1879}. More quantitatively, for hypothesis tests between such ``sub-Rayleigh" objects, 
the probability of correct identification
degrades as the PSF more severely perturbs the measured images~\cite{Goodman2005}.

A paradigm shift for sub-Rayeigh imaging recently emerged via the calculation of task-specific error bounds that optimize over all measurements permitted by quantum mechanics~\cite{Tsang2020}.
These ``quantum limits" revealed that direct measurements of the image-plane optical intensity profile are to blame for the catastrophic degree of error implied by the Rayleigh criterion, whereas alternative measurements yield far lower error than direct imaging for many tasks \cite{Tsang2016a,Dutton2019,Tsang2019,Zhou2019,Lupo2020}. Quantum limits, and ``quantum-optimal" measurements that achieve them, were found for specific hypothesis tests including one-vs-two point source discrimination \cite{Lu2018,Zhang2020b}
and exoplanet detection \cite{Huang2021,Zanforlin2022}. However, no general results exist that broadly apply to real-world object discrimination settings.

This Letter finds quantum limits and quantum-optimal measurements for generalized sub-Rayleigh object discrimination, 
with wide applicability to sub-cellular fluorescence microscopy, exoplanet surveys, pattern recognition in remote sensing, passive human iris identification, and many more imaging domains. 
For sub-Rayleigh hypothesis tests between \emph{any} two incoherent, quasi-monochromatic 2D objects, we 1)~compute the quantum Chernoff bound on asymptotic discrimination error, 2)~compute the classical Chernoff exponent that characterizes the error of with ideal direct imaging, 3)~quantify a quadratic scaling gap between the two Chernoff exponents, and 4)~identify a quantum-optimal measurement that employs a pre-detection spatial-mode sorting device whose linear-optical design does not depend on the object models. Remarkably, our results extend to $M$-ary discrimination: the same object-independent measurement is quantum-optimal for \emph{any} database of $M>2$ objects. Last, we define Hamming-like distance measures of object databases to quantify the realizable advantage over direct imaging.

\emph{Quantum model}---Let $H_j$, $j\in[1,M]$, denote a hypothesis corresponding to one of $M$ candidate objects. Under $H_j$, the quantum state $\eta_j$ on Hilbert space $\mathcal{H}$ describes one temporal mode of a quasi-monochromatic optical field collected by an imaging system. Many natural thermal sources
exhibit a mean photon flux $\epsilon\ll1$ per temporal mode such that multi-photon detection within the optical coherence time is vanishingly rare~\cite{Mandel1959}. Using a weak-source Fock expansion $\eta_j=(1-\epsilon)\outerproduct{0}{0}+\epsilon\rho_j+O(\epsilon^2)$,
where $\outerproduct{0}{0}$ is the vacuum state, the state $\rho_j$ carries all of the spatial information about the object under $H_j$~\cite{Tsang2016a}. Since $\rho_j$ is a state of one photon over multiple orthogonal spatial modes, it can be mapped to the state of a single bosonic mode on a Hilbert space spanned by the Fock states of that mode~\cite{Tsang2016a}. We denote this Hilbert space $\mathcal{H}^{(1)}$. 

Let an imaging system with a 2D coherent PSF $\psi(\vec{x})$ relate object- and image-plane position vectors $\vec{x}_{\rm obj}=\{x_{\rm obj},y_{\rm obj}\}$ and $\vec{x}=\mu\vec{x}_{\rm obj}$ by the transverse magnification $\mu$. We model the spatial irradiance of the object under $H_j$ by a normalized radiant exitance profile $m_j(\vec{x}_{\rm obj})$. The state of the collected optical field on $\mathcal{H}^{(1)}$ is then~\cite{Tsang2017}
\begin{equation}
	\rho_j=\iint^{\infty}_{-\infty}\frac{1}{\mu^2}m_j\bigg(\frac{\vec{x}}{\mu}\bigg)\outerproduct{\psi_{\vec{x}}}{\psi_{\vec{x}}}d^2\vec{x},
	\label{eq:rhoj}
\end{equation}
where the pure state $\ket{\psi_{\vec{x}}}=\iint_{-\infty}^{\infty}\psi(\vec{a}-\vec{x})\ket{\vec{a}}d^2\vec{a}$ encodes the effect of the aperture and $\ket{\vec{x}}
$ is a single-photon eigenket at image-plane position $\vec{x}$~\cite{Tsang2016a}. In a basis of orthogonal vectors $\ket{\phi_m}=\iint_{-\infty}^{\infty}\phi_{m}(\vec{x})\ket{\vec{x}}d^2\vec{x}$ that span $\mathcal{H}^{(1)}$, where $\phi_m(\vec{x})$ are orthogonal 2D functions,
the density matrix
\begin{equation}
	\rho_j=\sum_{m,n=0}^{\infty}d_{j,m,n}\outerproduct{\phi_{m}}{\phi_{n}}
	\label{eq:rhojPAD}
\end{equation}
has elements
$d_{j,m,n}=\iint_{-\infty}^{\infty}\mu^{-2}m_j(\vec{x}/\mu)c_{m,n}(\vec{x})d^2\vec{x}$, where $c_{m,n}(\vec{x})=\innerproduct*{\phi_{m}}{\psi_{\vec{x}}}\innerproduct*{\psi_{\vec{x}}}{\phi_{n}}$.

\begin{figure}
	\centering
	\includegraphics[width=1\columnwidth]{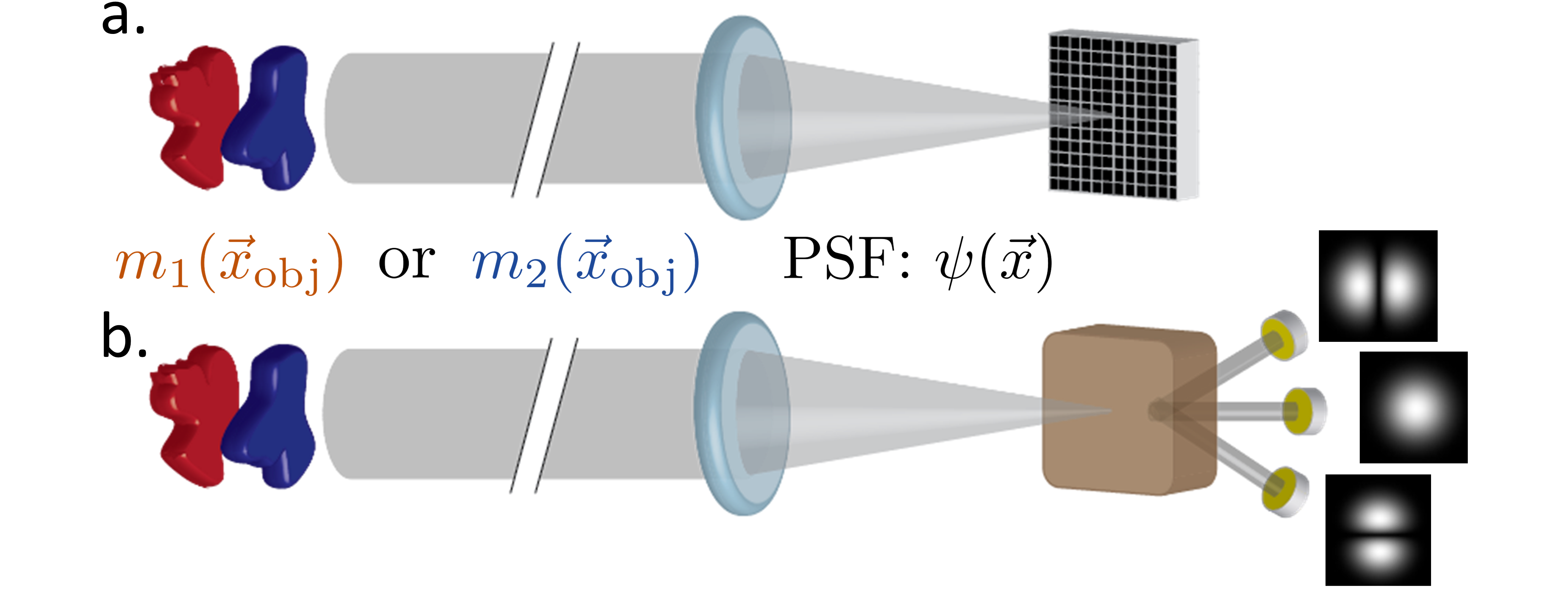}
	\vspace{-8pt}
	\caption{Discrimination of two objects $m_1(\vec{x}_{\rm obj})$ and $m_2(\vec{x}_{\rm obj})$. a.~Direct imaging. b.~TriSPADE receiver using a spatial-mode sorter and three shot-noise-limited photon detectors. For a Gaussian PSF $\psi(\vec{x})$, the sorted modes are shown at right. 
	}
	\label{fig:diagram}
\end{figure}

\emph{Quantum and classical detection theory}---Consider a hypothesis test between objects $m_1(\vec{x}_{\rm obj})$ and $m_2(\vec{x}_{\rm obj})$ with equal prior probabilities (Fig.~\ref{fig:diagram}). To make a decision $Z\in[1,2]$, a receiver measures the state $\eta_1^{\otimes \mathcal{M}}$ or $\eta_2^{\otimes \mathcal{M}}$ acquired over $\mathcal{M}$ temporal modes and then applies a pre-determined decision rule on the outcome(s). If the conditional probability of deciding $H_{j'}$ under true hypothesis $H_j$ is $P_{\mathcal{M}}(Z=j'\vert H_j)$, the average error probability $P_{\textrm{err},\mathcal{M}}=\big[P_{\mathcal{M}}(Z=1\vert H_2)+P_{\mathcal{M}}(Z=2\vert H_1)\big]/2$ is a symmetric performance metric for that measurement along with the decision rule~\footnote{While we only addresses symmetric hypothesis tests here, our perturbation theory for quantum information~\cite{Grace2021} will enable generalized results on asymmetric object discrimination via the quantum Stein lemma in future work.}. 
Optimizing over all such schemes, the quantum-limited minimum average error $P_{\textrm{err,min},\mathcal{M}}\sim e^{-\xi_{\textrm{Q}} \mathcal{M}}$ follows an exponential decay when $\mathcal{M}\gg1$, where the quantum Chernoff exponent (QCE) $\xi_{\rm Q}$ quantifies how efficiently each additional copy of the received state $\eta_j$ suppresses the minimum error~\cite{Audenaert2007,Nussbaum2009}. In~\cite{Supplemental}, we show that equivalently $P_{\textrm{err,min},\mathcal{M}}\sim e^{-\xi_{\rm Q}^{(1)}N}$, where $N=\epsilon \mathcal{M}$ is the average photon number of $\eta_j^{\otimes \mathcal{M}}$ and where the per-photon QCE~\cite{Audenaert2007,Nussbaum2009}
\begin{equation}
	\xi_{\textrm{Q}}^{(1)}=-\log\left[\min_{0\leq s \leq 1} \Tr\big(\rho_1^s\rho_2^{1-s}\big)\right]
	\label{eq:QuantumChernoff}
\end{equation}
obeys $\xi_{\rm Q}\approx\epsilon\xi_{\rm Q}^{(1)}$ for weak-source sub-Rayleigh objects.

The most general description of a measurement, a positive operator-valued measure (POVM), consists of a set of positive semi-definite operators $\{\Pi_z\}_{\mathcal{Z}}$ on $\mathcal{H}$, linked to measurement outcomes $\{z\}$ on an outcome space $\mathcal{Z}$, that resolve the identity operator as $\sum_{z\in\mathcal{Z}}\Pi_z=\mathcal{I}$~\cite{Helstrom1976}. For a particular measurement performed on $\eta_j^{\otimes\mathcal{M}}$, the minimum average error among all decision rules goes as $P_{\textrm{err,min,Meas},\mathcal{M}}\sim e^{-\xi_{\textrm{Meas}} \mathcal{M}}$, where $\xi_{\textrm{Meas}}$ is the Chernoff exponent (CE) for the chosen measurement~\cite{VanTrees2013,Yu2021}. For weak sources, we show~\cite{Supplemental} that the minimal error of any measurement that uses temporally-resolved photon counting goes as $P_{\textrm{err,min,Meas},\mathcal{M}}\sim e^{-\xi^{(1)}_{\textrm{Meas}} N}$, where $\xi_{\rm Meas}\approx\epsilon\xi^{(1)}_{\rm Meas}$ in the sub-Rayleigh regime and where~\cite{VanTrees2013}
\begin{equation}
	\xi_{\rm Meas}^{(1)}=-\log\Bigg[\min_{0\leq s \leq 1} \sum_{z\in\mathcal{Z}^{(1)}}P(z\vert\rho_1)^s P(z\vert\rho_2)^{1-s}\Bigg]
	\label{eq:Chernoff}
\end{equation} 
is the per-photon CE, which depends on probabilities $P(z\vert\rho_j)=\Tr(\Pi^{(1)}_z\rho_j)$ of outcomes in a single-photon subspace $\mathcal{Z}^{(1)}$ obtained by the reduced POVM $\{\Pi_z^{(1)}\}_{\mathcal{Z}^{(1)}}$ on $\mathcal{H}^{(1)}$.

The quantum and classical statistics are related by the quantum Chernoff bound $\xi_{\rm Meas}\leq\xi_{\rm Q}$; that is, the QCE automatically optimizes over the CEs of all POVMs on $\mathcal{H}^{\otimes \mathcal{M}}$~\footnote{The QCE optimizes over all POVMs including measurements that act collectively on multiple copies of the state $\eta_j$. For the present context, our results show that individual measurements on each copy of $\eta_j$ are sufficient to saturate the quantum Chernoff bound.}. A measurement whose per-photon CE matches the QCE ($\xi_{\rm Meas}^{(1)}=\xi_{\rm Q}^{(1)}$) is quantum-optimal for the given hypothesis test. Conversely, a gap ($\xi_{\rm Meas}^{(1)}<\xi_{\rm Q}^{(1)}$) indicates a fundamental sub-optimality in the measurement that cannot be remedied by data post-processing. 

\emph{Results: binary object discrimination}---In this section we compute the QCE $\xi_{\rm Q}^{(1)}$ for generalized sub-Rayleigh object discrimination and find a universally optimal measurement for which $\xi^{(1)}_{\rm Meas}=\xi^{(1)}_{\rm Q}$. For a preliminary result, 
if the object under $H_1$ is a single point source at the object-plane position $\vec{x}_{1, \rm obj}=\vec{x}_1/\mu$~\cite{Lu2018,Zhang2020b,Huang2021,Zanforlin2022}, 
we find that the QCE is exactly~\cite{Supplemental}
\begin{equation}
	\xi^{(1)}_{\textrm{Q}}\!=\!-\log\!\left[ \iint_{-\infty}^{\infty}\frac{1}{\mu^2}m_2\bigg(\frac{\vec{x}-\vec{x}_1}{\mu}\bigg)\abs{\Gamma(\vec{x})}^2d^2\vec{x}\right]\!,
	\label{eq:quantum_Chernoff_exact}
\end{equation}
where $\Gamma(\vec{x})=\innerproduct{\psi_{\vec{\Omega}}}{\psi_{\vec{x}}}$ is the 2D autocorrelation function of the PSF and $\vec{\Omega}$ denotes the origin of the image-plane coordinate system. In this case, $\xi_{\rm BSPADE}^{(1)}=\xi_{\rm Q}^{(1)}$ is achieved by a 2D binary spatial mode demultiplexing (BSPADE) device \cite{Tsang2016a,Boucher2020,Ang2017} that passively couples a PSF-matched spatial mode to one shot-noise-limited photon-counting detector (i.e., $\Pi_0=\outerproduct{\psi_{\vec{x}_1}}{\psi_{\vec{x}_1}}$) and all other light to a second detector (i.e., $\Pi_1=\mathcal{I}-\outerproduct{\psi_{\vec{x}_1}}{\psi_{\vec{x}_1}}$)~\cite{Supplemental}. As an example, for discriminating one-vs-two point sources with a 2D Gaussian PSF $\psi(\vec{x})=(2\pi \sigma^2)^{-1/2}\exp(-(x^2+y^2)/4\sigma^2)$, where $d$ is the source separation under $H_2$~\cite{Lu2018}, we confirm that the BSPADE CE enjoys a quadratic ($d^2$) scaling advantage  over the CE of idealized 2D direct imaging (an infinite spatial bandwidth, unity fill factor, unity quantum efficiency photon-counting detector array, Fig.~\ref{fig:diagram}a.) as $d\ll\sigma$~\cite{Supplemental}.

\begin{figure}[t]
	\centering
	\includegraphics[width=\columnwidth]{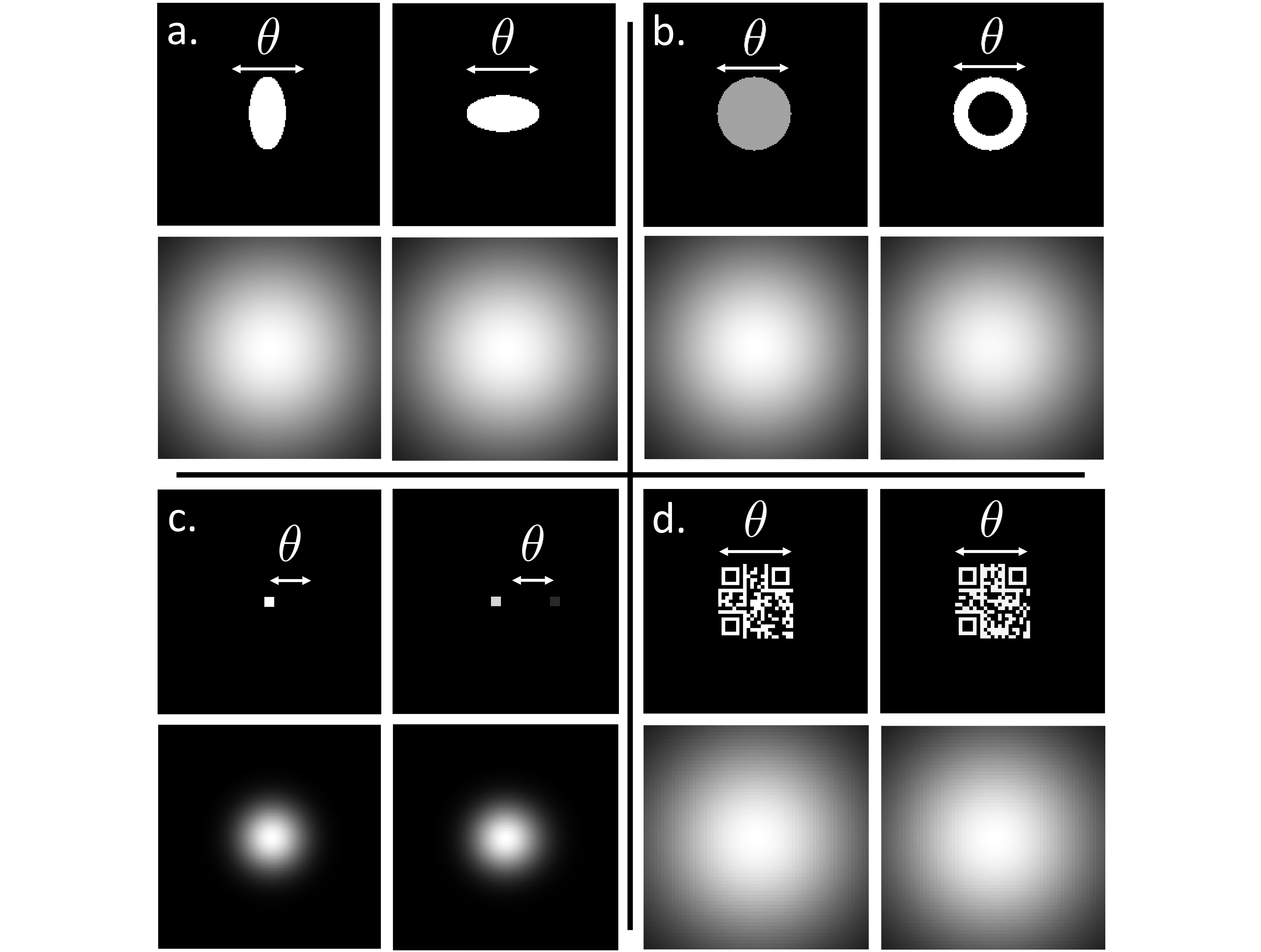}
	\caption{Simplified object pair examples: a.~vertical vs. horizontal ellipse, b.~filled vs. hollow nuclear pore~\cite{Thevathasan2019}, c.~exoplanet detection, d.~QR code reading. Upper images: normalized ground truth object irradiance. Lower images: Gaussian-PSF-convolved image-plane intensity profiles when $\gamma=1$.}
	\label{fig:Shapes}
\end{figure}

We now generalize to two arbitrary objects $m_1(\vec{x}_{\rm obj})$ and $m_2(\vec{x}_{\rm obj})$, with applications in bioimaging, astronomy, and computer vision (Fig.~\ref{fig:Shapes}). We focus on the sub-Rayleigh limit $\gamma\ll1$, where $\gamma=\mu\theta/\sigma$ quantifies the geometric ratio between the spatial extent of the objects~($\theta$) and the PSF width~($\sigma$).
We also define $\tilde{m}_j(\vec{x}_{\rm obj})=\theta^2 m_j(\theta\vec{x}_{\rm obj})$, $\tilde{\psi}(\vec{x})=\sigma \psi(\sigma\vec{x})$, and $\tilde{\Gamma}(\vec{x})=\Gamma(\sigma\vec{x})$ 
as non-dimensionalized representations of the objects, the PSF, and the PSF autocorrelation function, respectively, to isolate the effect of diffraction (i.e., $\gamma$) from that of the object and aperture~\cite{Supplemental}. 
We require that the objects' 2D centroids coincide at a location
known to the receiver from prior knowledge and/or a preliminary measurement \cite{Sajjad2021a,Grace2020c,deAlmeida2021}, such that the task is object identification and not localization, and that the PSF $\psi(\vec{x})$ is even in $x$ and $y$, as with a circularly symmetric aperture.

To derive the generalized QCE~\cite{Supplemental}, we represent $\rho_1$ and $\rho_2$ [Eq.~\eqref{eq:rhojPAD}] in a basis of PSF-adapted (PAD) eigenvectors $\ket{\phi_m}$ via Gram-Schmidt orthogonalization of the 2D Cartesian derivatives of the PSF $\psi(\vec{x})$~\cite{Kerviche2017a,Rehacek2017b,Tsang2018a}.
For a 2D Gaussian PSF, the PAD basis functions $\phi_m(\vec{x})$ are Hermite-Gauss polynomials \cite{Rehacek2017b}. After expanding $\rho_1$ and $\rho_2$ in powers of $\gamma\ll1$ and truncating to finite dimensions~\cite{Dutton2019}, we use operator perturbation theory~\cite{Grace2021} to find
\begin{equation}
	\begin{aligned}
		&\xi_{\rm Q}^{(1)}=\max_{0\leq s \leq 1}\!\big[\big(sm_{1,x^2}\!+\!(1\!-\!s)m_{2,x^2}\!-\!m_{1,x^2}^sm_{2,x^2}^{1\!-\!s}\big)\Gamma_{x^2}\\
		&+\!\big(sm_{1,y^2}\!+\!(1-s)m_{2,y^2}\!-\!m_{1,y^2}^sm_{2,y^2}^{1-s}\big)\Gamma_{y^2}\big]\gamma^2\!+\!O(\gamma^3),
	\end{aligned}
	\label{eq:GeneralQCE}
\end{equation}
where $m_{j,x^ky^l}=\iint_{-\infty}^{\infty}x_{\rm obj}^ky_{\rm obj}^l\tilde{m}_j(\vec{x}_{\rm obj})d^2\vec{x}_{\rm obj}$ are spatial moments of the non-dimensionalized object models and $\Gamma_{x^ky^l}= -[\textrm{Re}(\partial^{k+l}\tilde{\Gamma}(\vec{x})/\partial x^k\partial y^l)]_{\vec{x}=\vec{\Omega}}$ are derivatives of the PSF autocorrelation function. The QCE in Eq.~\eqref{eq:GeneralQCE} is our first main result and represents the quantum limit for discrimination between \textit{any} two incoherent objects in the sub-Rayleigh limit $\gamma\ll1$. 

We compute the CE for direct imaging with a zeroless PSF~\footnote{While Eq.~\eqref{eq:CEDirectGeneralSeparable} does not hold for PSFs that are zero-valued at one or more locations, the direct imaging CE still retains the scaling $\xi_{\rm Direct}^{(1)}\sim \gamma^4$~\cite{Supplemental}.} that is separable in $x$ and $y$
to be~\cite{Supplemental}
\begin{equation}
	\xi^{(1)}_{\textrm{Direct}}=(1/32)(\mathcal{K}_x+\mathcal{K}_y)\gamma^4+O(\gamma^5),
	\label{eq:CEDirectGeneralSeparable}
\end{equation}
with $\mathcal{K}_a=(m_{1,a^2}-m_{2,a^2})^2\iint_{-\infty}^{\infty}\psi_{a^2}(\vec{x})^2/\abs{\tilde{\psi}(\vec{x})}^{2}d^2\vec{x}$ for $a\in[x,y]$
and where $\psi_{x^ky^l}(\vec{x})=\partial^{k+l}\abs{\tilde{\psi}(\vec{x})}^2/\partial x^k \partial y^l$ are derivatives of the incoherent PSF. Eqs.~\eqref{eq:GeneralQCE} and \eqref{eq:CEDirectGeneralSeparable} reveal a quadratic scaling sub-optimality in direct imaging---$\xi^{(1)}_{\rm Direct}\sim \gamma^4$ vs $\xi^{(1)}_{\rm Q}\sim \gamma^2$---for \emph{all} binary discrimination tasks~\footnote{In the Supplemental Material we analyze caveats to these results regarding PSFs with zeros as well as object pairs that have either identical second moments or different first moments. These special cases maintain the qualitative features of our results while indicating a general trend that the relative performance gap between $\xi_{\rm Q}^{(1)}$ and $\xi_{\rm Direct}^{(1)}$ generally increases for more demanding discrimination tasks.}. Alternatively, we introduce a ``TriSPADE" measurement (Fig.~\ref{fig:diagram}b.) that sorts the collected light between the PSF-matched spatial mode and the first-order PAD-basis modes in two perpendicular dimensions. TriSPADE uses only linear optics and shot-noise limited photodetectors to implement a measurement with projectors $\Pi_0=\outerproduct{\phi_0}{\phi_0}$, $\Pi_1=\outerproduct{\phi_1}{\phi_1}$, and  $\Pi_2=\outerproduct{\phi_2}{\phi_2}$ that do not depend on the candidate object models. The resulting CE $\xi^{(1)}_{\rm TriSPADE}$ achieves the QCE when $\gamma\ll1$~\cite{Supplemental}, meaning that TriSPADE is a quantum-optimal measurement for binary sub-Rayleigh object discrimination.

\begin{figure}[t]
	\centering
	\includegraphics[width=.99\columnwidth]{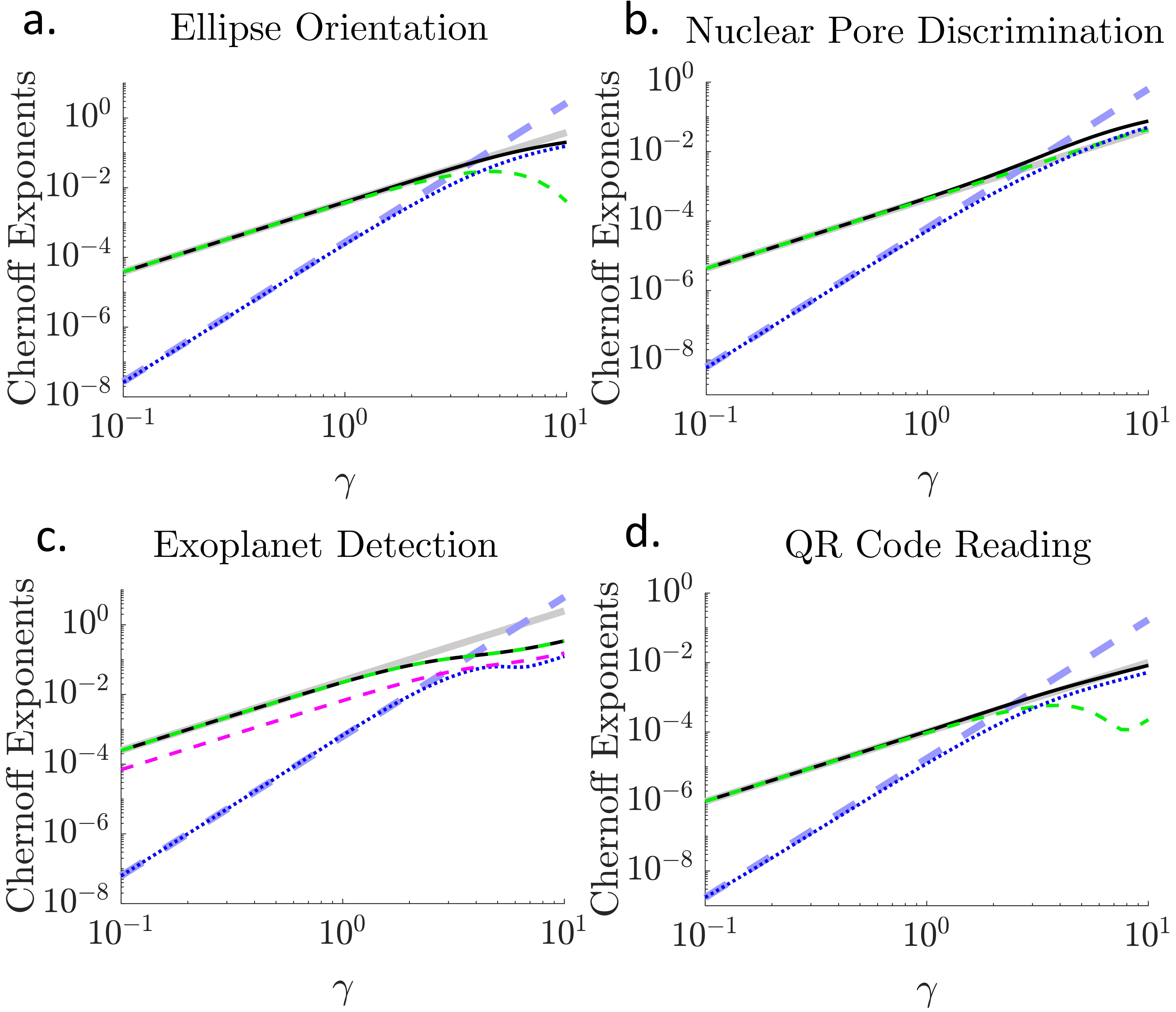}
	\caption{QCEs and CEs for the tasks from Fig.~\ref{fig:Shapes} with a 2D Gaussian PSF. Thick lines: analytical lowest-order (in $\gamma$) results for $\xi^{(1)}_{\rm Q}$ (solid) and $\xi^{(1)}_{\rm Direct}$ (dashed). Thin lines: numerical results for $\xi^{(1)}_{\rm Q}$ (solid), $\xi^{(1)}_{\rm Direct}$ (dotted), and $\xi^{(1)}_{\rm TriSPADE}$ (dashed). A misalignment of $\theta/10$ is used for the lower TriSPADE CE (magenta dashed, color online) in c.}
	\label{fig:ChernoffExponents}
\end{figure}

To illustrate these results, in Fig.~\ref{fig:ChernoffExponents} we numerically evaluate $\xi^{(1)}_{\rm Q}$, $\xi^{(1)}_{\rm Direct}$, and $\xi^{(1)}_{\rm TriSPADE}$ for the examples depicted in Fig.~\ref{fig:Shapes}.
The lowest-order behavior of the QCE in $\gamma$ [Eq.~\eqref{eq:GeneralQCE}] is an excellent approximation for both the full QCE and the TriSPADE CE throughout the sub-Rayleigh regime ($\gamma<1$), and the direct imaging results clearly exhibit the expected $O(\gamma^2)$ scaling gap.  
We also find TriSPADE to be robust to optical misalignment; a mode sorter that is misaligned from the object centroid retains the quadratic scaling advantage over direct imaging (Fig.~\ref{fig:ChernoffExponents}c). These results suggest that TriSPADE can perform a wide range of sub-Rayleigh hypothesis tests with substantially less error than conventional methods.

\emph{Results: $M$-ary object discrimination}---We now extend our analysis to $M>2$ equiprobable objects, such as a database of QR codes (Fig.~\ref{fig:Shapes}d.). The $M$-ary QCE $\xi_{\textrm{Q},M}^{(1)}=\min_{i\neq j}\xi_{\textrm{Q},i,j}^{(1)}$, which characterizes the quantum-limited asymptotic error for discriminating $M$ states, is found by minimizing the pairwise QCEs $\xi_{\textrm{Q},i,j}^{(1)}$ for each pair of states $\{\rho_i,\rho_j\}$~\cite{Li2016}. The similarly defined $M$-ary CE $\xi_{\textrm{Meas},M}^{(1)}=\min_{i\neq j}\xi_{\textrm{Meas},i,j}^{(1)}$ obeys the multiple quantum Chernoff bound $\xi_{\textrm{Meas},M}^{(1)}\leq\xi_{\textrm{Q},M}^{(1)}$~\cite{Li2016}. We have shown that TriSPADE saturates the pairwise quantum limit ($\xi_{\textrm{TriSPADE},i,j}^{(1)}=\xi_{\textrm{Q},i,j}^{(1)}$) for any two states when $\gamma\ll1$. Therefore, the TriSPADE measurement, which crucially does not depend on the candidate states, will always \emph{simultaneously} achieve the QCE for all pairs of states in a database. It follows that TriSPADE saturates the multiple quantum Chernoff bound with equality (i.e.,  $\xi_{\textrm{TriSPADE},M}^{(1)}=\xi_{\textrm{Q},M}^{(1)}$) when $\gamma\ll1$, so TriSPADE is a quantum-optimal measurement for discriminating among \emph{any} $M$-object database in the sub-Rayleigh limit.

\begin{figure}[t]
	\centering
	\includegraphics[width=1\columnwidth]{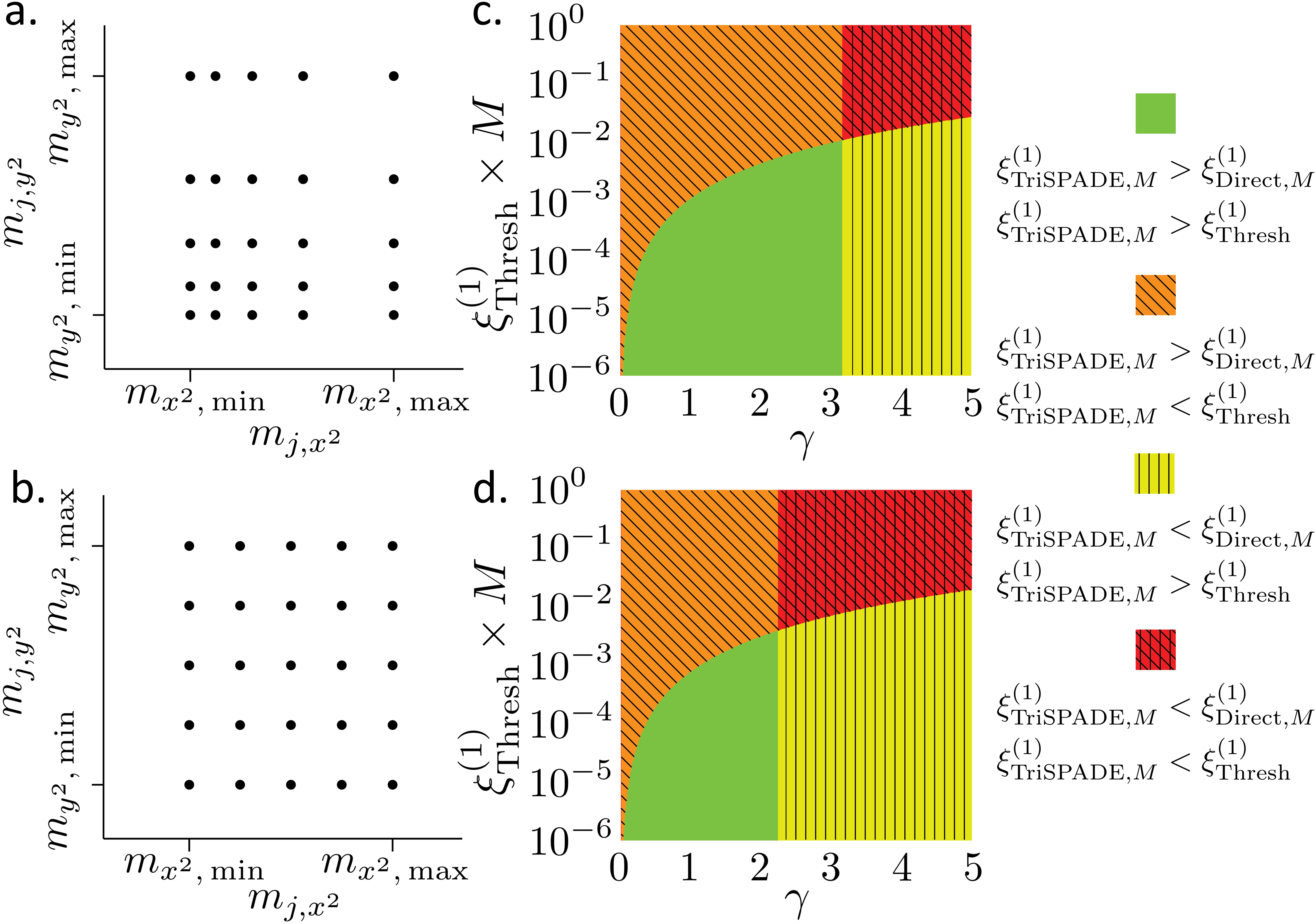}
	\caption{(a.-b.) Visualization of generalized object databases with quadratically and linearly packed 2D second moments. (c.-d.) Comparison of the CEs for TriSPADE, direct imaging and an error threshold for $M$-ary object discrimination to lowest order in $\gamma$ ($m_{x^2,\rm min}=0.05$, $m_{x^2,\rm max}=0.1$, color online).} 
\vspace{-15pt}
\label{fig:regionsHatches}
\end{figure}

Furthermore, we identify ``distance" measures that determine the constant-factor behavior of $\xi_{\textrm{Q},M}^{(1)}\sim\gamma^2$ and $\xi_{\textrm{Direct},M}^{(1)}\sim\gamma^4$ from the minimum distance among all object pairs in  a database. Our measures, which depend on relative second moments ($m_{j,x^2}$ and $m_{j,y^2}$) between candidate object models, resemble the Hamming distance in linear coding theory, which quantifies the distinguishability of noise-corrupted codewords~\cite{Hamming1950}. Approximating Eq.~\eqref{eq:GeneralQCE} using the quantum Bhattacharyya bound \cite{Pirandola2008}, we find that relative square roots of object second moments (e.g., $\sqrt{m_{i,x^2}}-\sqrt{m_{j,x^2}}$) form a distance for the pairwise QCEs, such that $M=M_xM_y$ quadratically packed objects on a $M_x\times M_y$ rectangular grid within the 2D space of second moments (Fig.~\ref{fig:regionsHatches}a.) forms an equidistant database in the sense that nearest neighbor objects along either the $x$ or $y$ direction all share equivalent pairwise exponents $\xi_{\textrm{Q},i,j}^{(1)}$ and $\xi_{\textrm{Direct},i,j}^{(1)}$. For an equidistant database of objects with second moments constrained between $m_{y^2,\rm min}=m_{x^2,\rm min}$ and $m_{y^2,\rm max}=m_{x^2,\rm max}$, we find~\cite{Supplemental}
\begin{align}
\label{eq:QCE_M_quadratic} \xi_{\textrm{Q},M}^{(1)}\!\approx&\frac{(\sqrt{m_{x^2,\rm max}}-\sqrt{m_{x^2,\rm min}})^2\Gamma_{x^2}}{2(M_{x}-1)^2}\gamma^2\!+\! O\big(\gamma^3\big),\\
\xi_{\textrm{Direct},M}^{(1)}\!=&\frac{(\sqrt{m_{x^2,\rm max}}-\!\sqrt{m_{x^2,\rm min}})^2\Psi_{x^2}}{8(M_{x}-1)^2m_{x^2,\rm min}^{-1}}\gamma^4\!+\! O\big(\gamma^5\big),
\label{eq:CE_M_quadratic} 
\end{align}
when $M\gg 1$, where the $x$ direction is chosen to minimize the distance and $\Psi_{x^2}=\iint_{-\infty}^{\infty}\psi_{x^2}(\vec{x})^2/\abs{\tilde{\psi}(\vec{x})}^2 d^2\vec{x}$. For direct imaging, the differences  $m_{i,x^2}-m_{j,x^2}$ constitute a distance measure for $\xi_{\textrm{Direct},M}^{(1)}$ [Eq.~\eqref{eq:CEDirectGeneralSeparable}], so linearly packed objects form an equidistant database (Fig.~\ref{fig:regionsHatches}b.). In this case, when $M\gg1$ we find~\cite{Supplemental}
\begin{align}
\label{eq:QCE_M_linear} \xi_{\textrm{Q},M}^{(1)}\!\approx&\frac{(m_{x^2,\rm max}-m_{x^2,\rm min})^2\Gamma_{x^2}}{8(M_{x}-1)^2m_{x^2,\rm max}}\gamma^2\!+O\big(\gamma^3\big),\\
\label{eq:CE_M_linear} \xi_{\textrm{Direct},M}^{(1)}\!=&\frac{(m_{x^2,\rm max}-m_{x^2,\rm min})^2\Psi_{x^2}}{32(M_{x}-1)^2}\gamma^4\!+O\big(\gamma^5\big).
\end{align}

To unravel the role of second-moment distances on generalized $M$-ary discrimination performance, we first probe the conditions under which the quantum-optimal TriSPADE receiver achieves a useful performance gain over conventional imaging. In Fig.~\ref{fig:regionsHatches}c.-d., we specify a 2D Gaussian aperture and depict parameterized regions for which, to lowest order in $\gamma$, TriSPADE attains a relative advantage over direct imaging (i.e., $\xi_{\textrm{TriSPADE},M}^{(1)}>\xi_{\textrm{Direct},M}^{(1)}$) and/or over a threshold representing an acceptable application-specific error rate (i.e., $\xi_{\textrm{TriSPADE},M}^{(1)}>\xi_{\rm Thresh}^{(1)}$). For both quadratic and linear databases with $M_y=M_x$, TriSPADE outperforms direct imaging only for sufficiently sub-Rayleigh objects: the region of TriSPADE's useful advantage over direct imaging is bounded by $\gamma<1/\sqrt{2m_{x^2,\rm min}}$ with quadratic packing, whereas with linear packing it is bounded by the tighter  $\gamma<1/\sqrt{2m_{x^2,\rm max}}$.

\begin{figure}[t]
\centering
\includegraphics[width=.98\columnwidth]{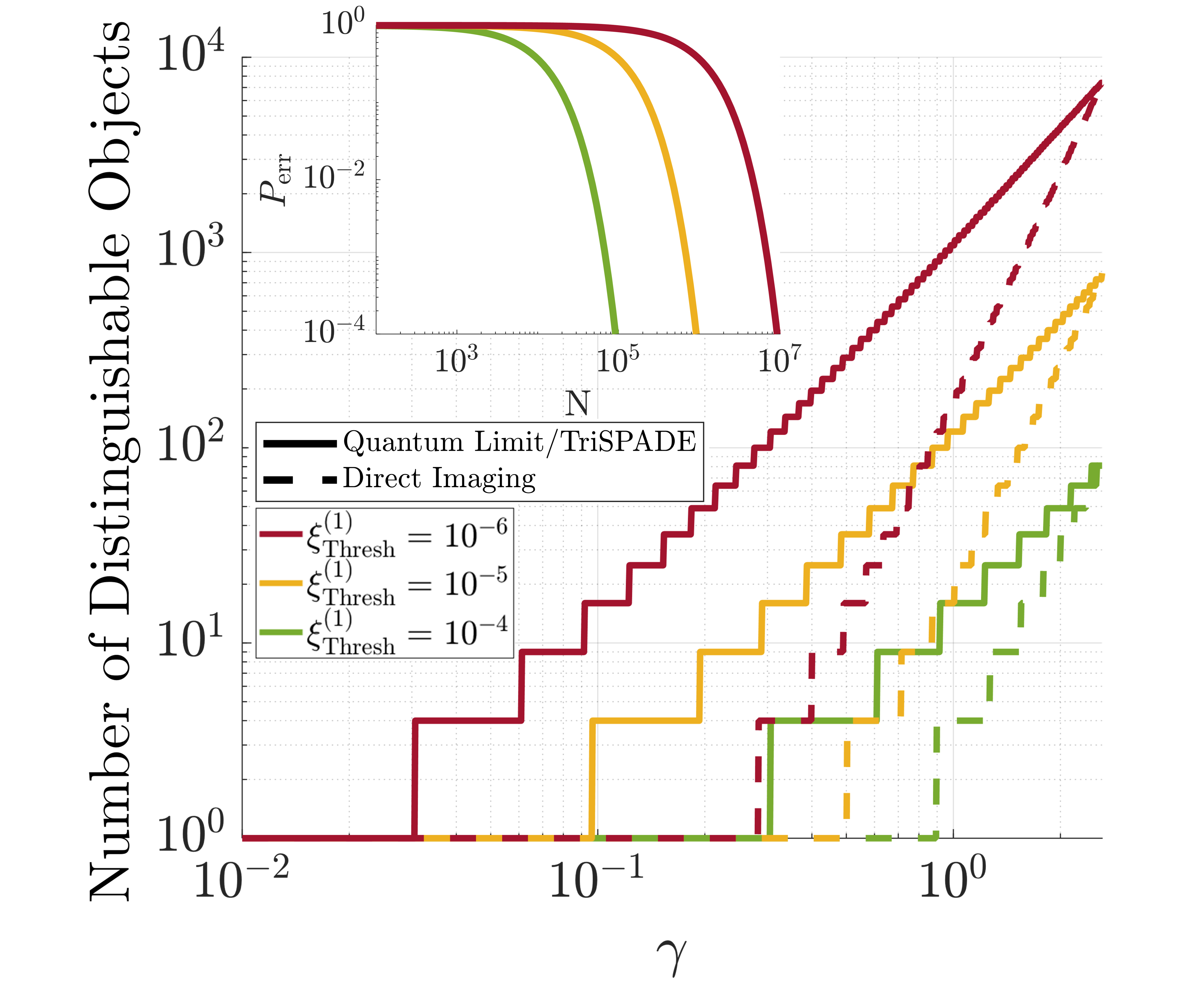}
\caption{Maximum number of objects that are distinguishable at a threshold error rate to lowest order in $\gamma$ with a 2D Gaussian aperture ($m_{x^2,\rm min}=0.05$, $m_{x^2,\rm max}=0.1$, color online). Inset: error probability vs. mean detected photon number.}
\label{fig:ResolvableObjects}
\end{figure}

Finally, in Fig.~\ref{fig:ResolvableObjects} we ask how many objects can be distinguished to a desired accuracy with a conventional or quantum-optimal measurement, which directly relates the the decision-making power of an autonomous imaging system, for example. 
Using an equidistant database for both cases, we solve Eqs.~\eqref{eq:QCE_M_quadratic} and \eqref{eq:CE_M_linear} for $M=M_{x}^2$ to lowest order in $\gamma$ and find that TriSPADE resolves more objects than direct imaging when $\gamma<\sqrt{2}/(\sqrt{m_{x^2,\rm max}}+\sqrt{m_{x^2,\rm min}})$ regardless of the threshold error rate $\xi_{\rm Thresh}^{(1)}$. As the threshold is relaxed, meaning more photons are available and/or more error can be tolerated (inset), the gap between TriSPADE and direct imaging grows to over two orders of magnitude for small $\gamma$. We conclude that TriSPADE significantly increases the complexity of distinguishable sub-Rayleigh object databases without compromising performance.


\emph{Conclusion}---Our work shows that a realizable optical receiver could lead to substantial improvements in decision-making based on super-resolution biological, astronomical, and terrestrial imaging. 
System imperfections, such as optical losses, mode crosstalk, or detector noise, can only reduce discrimination accuracy from the ideal case. Our quantum limit calculation will therefore be useful in definitively ruling out quantitative regimes of discrimination capability for future imaging systems.

\begin{acknowledgments}
	The authors acknowledge useful discussions with Amit Ashok, Mark Neifeld and Jeff Shapiro on the results and the manuscript. This research was supported by the DARPA IAMBIC Program under contract number HR00112090128. The views, opinions and/or findings expressed are those of the authors and should not be interpreted as representing the official views or policies of the Department of Defense or the U.S. Government.
\end{acknowledgments}

\bibliography{Quantum_Hypothesis_Testing-Sub-Diffraction_Quantum_Hypothesis_Testing}

\newpage
\section{Supplemental Material}
\subsection{Per-Photon (Quantum) Chernoff Exponents}
For a binary hypothesis test between two states $\eta_1$ and $\eta_2$ on $\mathcal{H}$, the QCE is defined as \cite{Audenaert2007,Nussbaum2009}
\begin{equation}
	\xi_{\rm Q}=-\log\left[\min_{0\leq s \leq 1} \Tr\big(\eta_1^s\eta_2^{1-s}\big)\right].
	\label{eq:QCE_eta}
\end{equation}
Expanding the states into their Fock-space components, i.e., $\eta_j=(1-\epsilon)\outerproduct{0}{0}+\epsilon\rho_j+O\big(\epsilon^2\big)$, we have
\begin{equation}
	\eta_j^s = (1-\epsilon)^s\outerproduct{0}{0}+\epsilon^s\rho_j^s+O\big(\epsilon^{2s}\big),
	\label{eq:eta_j^s}
\end{equation}
where each term can be individually exponentiated due to the orthogonality of the photon-number subspaces in the Fock decomposition and where $\outerproduct{0}{0}^s=\outerproduct{0}{0}$. Thus,
\begin{equation}
	\begin{aligned}
		\Tr\big(\eta_1^s\eta_2^{1-s}\big)&=\Tr\Big\{\big[(1-\epsilon)^s\outerproduct{0}{0}+\epsilon^s\rho_1^s+O\big(\epsilon^{2s}\big)\big]\\
		&\!\!\!\!\!\!\!\!\!\!\!\!\!\!\!\!\!\!\!\!\!\!\!\!\!\!\!\!\!\times\big[(1-\epsilon)^{1-s}\outerproduct{0}{0}+\epsilon^{1-s}\rho_2^{(1-s)}+O\big(\epsilon^{2(1-s)}\big)\big]\Big\}\\
		&=1-\epsilon+\epsilon\Tr\big(\rho_1^s\rho_2^{1-s}\big)+\Tr\big[O\big(\epsilon^2\big)\big],
	\end{aligned}
	\label{eq:Treta1seta21-s}
\end{equation}
where orthogonality of the photon-number subspaces is used for the second equality (i.e., $\mel{0}{\rho_j}{0}=0$). Including the minimization over $s$ and using the definitions of $\xi_{\rm Q}$ and $\xi_{\rm Q}^{(1)}$ in Eq.~\eqref{eq:QuantumChernoff} from the main text and Eq.~\eqref{eq:QCE_eta}, we rearrange terms to find
\begin{equation}
	1-e^{-\xi_{\rm Q}}=\epsilon\left(1-e^{-\xi^{(1)}_{\rm Q}}\right)+O\big(\epsilon^2\big).
	\label{eq:xiQ_exponential_relation}
\end{equation}
With deeply sub-Rayleigh objects, the regime of focus for our results, the dominating effect of diffraction causes the collected optical field from each of the two objects to be nearly identical. Thus, the states $\eta_1$ and $\eta_2$ converge to one another such that $\xi_{\rm Q}\approx 0$; similarly, $\rho_1$ and $\rho_2$ are nearly identical such that $\xi_{\rm Q}^{(1)}\approx 0$. In this case, under the weak-source approximation with $\epsilon\ll1$ we find the result $\xi_{\rm Q}\approx\epsilon\xi_{\rm Q}^{(1)}$.

For the same hypothesis test, the definition of the CE for a POVM $\{\Pi_z\}_{\mathcal{Z}}$ on $\mathcal{H}$ is~\cite{VanTrees2013}
\begin{equation}
	\xi_{\rm Meas}=-\log\left[\min_{0\leq s \leq 1} \sum_{z\in\mathcal{Z}}P(z\vert\eta_1)^s P(z\vert\eta_2)^{1-s}\right],
	\label{eq:CE_eta}
\end{equation}
where $P(z\vert\eta_j)=\Tr(\Pi_z\eta_j)$. For a measurement that performs photon counts in each individual temporal mode, the single-temporal-mode measurement outcome space $\mathcal{Z}$ can be decomposed into a countably infinite number of subspaces $\mathcal{Z}^{(n)}$, $n\in[0,\infty]$, each corresponding to the detection of $n$ photons in the given temporal mode. The CE therefore becomes 
\begin{equation}
	\xi_{\rm Meas}=-\log\left[\min_{0\leq s \leq 1} \sum_{n=0}^{\infty}\sum_{z\in\mathcal{Z}^{(n)}}P(z\vert\eta_1)^s P(z\vert\eta_2)^{1-s}\right].
	\label{eq:CE_eta_decomposed}
\end{equation}
Under the weak-source approximation, when multi-photon detection events can be neglected, the CE expression retains only the first two terms in the outermost sum corresponding to the $\mathcal{Z}^{(0)}$ and $\mathcal{Z}^{(1)}$ outcome subspaces. Clearly, the detection of zero photons can be considered as a single outcome, which will occur with probability $1-\epsilon$ under both hypotheses. The detection of one photon will occur within the set of outcomes $\mathcal{Z}^{(1)}$ and will depend only on the single-photon state $\rho_j$ and on the component of the POVM contained within the single-photon subspace of $\mathcal{H}$ subspace. Defining this truncated POVM on $\mathcal{H}^{(1)}$ as $\{\Pi_z^{(1)}\}_{\mathcal{Z}^{(1)}}$, the measurement outcome probabilities conditioned on detection of a single photon become $P(z\vert\rho_j)=\Tr(\Pi^{(1)}_z\rho_j)$. Since the probability of single photon detection within each temporal mode is $\epsilon$, for any $z\in\mathcal{Z}^{(1)}$ we have $P(z\vert\eta_j)=\epsilon P(z\vert{\rho_j})$, and thus the full CE becomes
\begin{equation}
	\begin{aligned}
		\xi_{\rm Meas}\approx&-\log\Bigg[1-\epsilon \\
		&+ \epsilon\min_{0\leq s \leq 1} \sum_{z\in\mathcal{Z}^{(1)}}P(z\vert\rho_1)^s P(z\vert\rho_2)^{1-s}\Bigg].
	\end{aligned}
	\label{eq:CE_eta_0and1}
\end{equation}
Using Eq.~\eqref{eq:Chernoff} from the main text and Eq.~\eqref{eq:CE_eta}, we find
\begin{equation}
	1-e^{-\xi_{\rm Meas}}\approx\epsilon\left(1-e^{-\xi^{(1)}_{\rm Meas}}\right).
	\label{eq:xiMeas_exponential_relation}
\end{equation}
Again, for sub-Rayleigh objects the measurement outcome probabilities will be nearly identical for the two hypotheses so that $\xi_{\rm Meas}\approx0$ and $\xi_{\rm Meas}^{(1)}\approx0$, and therefore $\xi_{\rm Meas}\approx\epsilon\xi_{\rm Meas}^{(1)}$.

\subsection{Point Source vs Arbitrary Object Discrimination}
Let the candidate objects be a point source at position $\vec{x}_{1, \rm obj}=\vec{x}_1/\mu$, described by a pure state $\rho_1=\ket{\psi_{\vec{x}_1}}\bra{\psi_{\vec{x}_1}}$, and a second object $m_2(\vec{x}_{\rm obj})$ with state $\rho_2$~(Eq.~\eqref{eq:rhoj} from the main text). For pure-vs-mixed state hypothesis tests, the projective measurement with POVM elements $\Pi_0=\rho_1$ and $\Pi_1= \mathcal{I}-\rho_1$ is known to be quantum-optimal, and the QCE takes the simplified form  $\xi_{\textrm{Q}}^{(1)}=-\log[\Tr(\rho_1\rho_2)]$~\cite{Kargin2005}. In the incoherent imaging context, this measurement can be performed with a 2D BSPADE aligned to the point source location $\vec{x}_{1,\rm obj}$ \cite{Tsang2016a,Boucher2020,Ang2017}, which implements the POVM with elements $\Pi_0=\outerproduct{\psi_{\vec{x}_1}}{\psi_{\vec{x}_1}}$ and $\Pi_1=\mathcal{I}-\outerproduct{\psi_{\vec{x}_1}}{\psi_{\vec{x}_1}}$. Assuming a shift invariant imaging system and paraxial optics~\cite{Goodman2005}, the distinguishability of the two objects will not be changed by a global coordinate shift such that $\rho_1=\outerproduct{\psi_{\vec{\Omega}}}{\psi_{\vec{\Omega}}}$. The shifted second object is then $m_2(\vec{x}_{\rm obj}-\vec{x}_{1, \rm obj})$. Using the shifted coordinates, the QCE canbe written exactly in the form of Eq.~\eqref{eq:quantum_Chernoff_exact} from the main text for any arbitrary object model $m_2(\vec{x}_{\rm obj})$. 

One special case is discrimination between a single point source, modeled by the state $\rho_1=\outerproduct{\psi_{\vec{\Omega}}}{\psi_{\vec{\Omega}}}$, and two point sources, modeled by $\rho_2=(1/2)(\outerproduct{\psi_{\{-d/2,0\}}}{\psi_{\{-d/2,0\}}}+\outerproduct{\psi_{\{d/2,0\}}}{\psi_{\{d/2,0\}}})$, where $d$ is the spatial separation between the two point sources along the $x$ axis. Using a 2D Gaussian PSF $\psi(\vec{x})=(2\pi \sigma^2)^{-1/2}\exp(-(x^2+y^2)/4\sigma^2)$, our result in Eq.~\eqref{eq:quantum_Chernoff_exact} with $\mu=1$ and $\sigma=1$ reproduces the reported QCE of $\xi_{\rm Q}^{(1)}=d^2/16$~\cite{Lu2018}, which is equaled by the CE of the quantum-optimal 2D BSPADE measurement. For comparison, typical receivers for incoherent imaging use the direct imaging measurement (Fig.~\ref{fig:diagram}a.), which consists of an image-plane detector array (e.g., a camera sensor) that records arrival times and positions of incident photons. Idealized 2D direct imaging can be described as a continuously sampled projective measurement with differential POVM elements $d\Pi_{\vec{x}}=\outerproduct{\vec{x}}{\vec{x}}d^2\vec{x}$, where $\int_{\mathbb{R}^2}d\Pi_{\vec{x}}=\mathcal{I}$.
Ref.~\cite{Lu2018} found that the CE for ideal direct imaging is $\xi_{\rm Direct}^{(1)}\approx d^4/256$ to lowest order in $d\ll1$, indicating a quadratic scaling gap in the achievable error exponent compared with the quantum limit as the point-source separation under $H_2$ is made small.

\subsection{Non-Dimensionalized Quantities: $\tilde{m}_j(\vec{x}_{\rm obj})$, $\tilde{\psi}(\vec{x})$, $\ket{\tilde{\psi}_{\vec{x}}}$,  $\tilde{\Gamma}(\vec{x})$, $\tilde{\phi}_m(\vec{x})$, $\ket{\tilde{\phi}_m}$, and $\tilde{c}_{m,n}(\vec{x})$}
We first justify the normalized definitions for the non-dimensionalized quantities $\tilde{m}_j(\vec{x}_{\rm obj})$, $\tilde{\psi}(\vec{x})$ and $\tilde{\Gamma}(\vec{x})$. We seek normalization factors $a$ and $b$ such that $\tilde{m}_j(\vec{x}_{\rm obj})=a m_j(\theta \vec{x}_{\rm obj})$ and $\tilde{\psi}(\vec{x})=b \psi(\sigma \vec{x})$ satisfy the normalization conditions $1=\iint_{-\infty}^{\infty}\tilde{m}_j(\vec{x}_{\rm obj})d^2\vec{x}_{\rm obj}$ and $1=\iint_{-\infty}^{\infty}\tilde{\psi}^\ast(\vec{x})\tilde{\psi}(\vec{x})d^2\vec{x}$. Making the coordinate transformations $\vec{x}_{\rm obj}'=\theta\vec{x}_{\rm obj}$ and $\vec{x}'=\sigma\vec{x}$, we have
\begin{equation}
	\begin{aligned}
		1=&\frac{a}{\theta^2}\iint_{-\infty}^{\infty}m_j(\vec{x}_{\rm obj}')d^2\vec{x}_{\rm obj}'\\
		1=&\frac{\abs{b}^2}{\sigma^2}\iint_{-\infty}^{\infty}\psi^\ast(\vec{x}')\psi(\vec{x}')d^2\vec{x}'.	
	\end{aligned}
	\label{eq:normalizations}
\end{equation}
Since each of the integrals in Eq. \eqref{eq:normalizations} is equal to unity, we confirm the normalizations $a=\theta^2$ and $b=\sigma$. Using the definition $\tilde{\Gamma}(\vec{x})=\Gamma(\sigma\vec{x})$, we then have
\begin{equation}
	\begin{aligned}
		\tilde{\Gamma}(\vec{x})=&\iint_{-\infty}^{\infty}\psi^\ast(\vec{a})\psi(\vec{a}-\sigma\vec{x})d^2\vec{a}\\
		=&\iint_{-\infty}^{\infty}\sigma^2\psi^\ast(\sigma\vec{a})\psi(\sigma\vec{a}-\sigma\vec{x})d^2\vec{a}\\
		=&\iint_{-\infty}^{\infty}\tilde{\psi}^\ast(\vec{a})\tilde{\psi}(\vec{a}-\vec{x})d^2\vec{a}\\
		=&\innerproduct{\tilde{\psi}_{\vec{\Omega}}}{\tilde{\psi}_{\vec{x}}},
	\end{aligned}
	\label{eq:GammaTilde}
\end{equation}
where the coordinate transformation $\vec{a}\to\sigma\vec{a}$ is made in the second equality, the definition of $\tilde{\psi}(\vec{x})$ is used for the third equality, and we define $\ket{\tilde{\psi}_{\vec{x}}}=\iint_{-\infty}^{\infty}\tilde{\psi}(\vec{x})\ket{\vec{x}}d^2\vec{x}$ for the final equality.

Likewise, given an arbitrary orthonormalized basis with eigenvectors $|\phi_m\rangle$, the normalization condition $\iint_{-\infty}^{\infty}\phi_{m}(\vec{x})^{\ast}\phi_n(\vec{x})d^2\vec{x}=\delta_{m,n}$ results in the non-dimensionalized basis functions $\tilde{\phi}_m(\vec{x})=\sigma \phi_m(\sigma\vec{x})$. Defining $\tilde{c}_{m,n}(\vec{x})=c_{m,n}(\sigma\vec{x})$, analogous steps to those in Eq.~\eqref{eq:GammaTilde} can be used to show that
\begin{equation}
	\begin{aligned}
		\tilde{c}_{m,n}(\vec{x})=&\iint_{-\infty}^{\infty}\tilde{\phi}_m^\ast(\vec{a})\tilde{\psi}(\vec{a}-\vec{x})d^2\vec{a}\\
		&\times\iint_{-\infty}^{\infty}\tilde{\psi}^\ast(\vec{a}-\vec{x})\tilde{\phi}_n(\vec{a})d^2\vec{a}\\
		=&\innerproduct{\tilde{\phi}_m}{\tilde{\psi}_{\vec{x}}}\innerproduct{\tilde{\psi}_{\vec{x}}}{\tilde{\phi}_n},
	\end{aligned}
	\label{eq:cmntilde}
\end{equation}
where $\ket{\tilde{\phi}_m}=\iint_{-\infty}^{\infty}\tilde{\phi}_{m}(\vec{x})\ket{\vec{x}}d^2\vec{x}$.

\subsection{The QCE for Binary Sub-Rayleigh Object Discrimination}
\begin{table}[b]
	\centering
	\begin{tabular}{|c|cccccc|}
		\hline
		$m$ & 0 & 1 & 2 & 3 & 4 & 5\\
		\hline
		$k_m$ & 0 & 1 & 0 & 2 & 1 & 0\\
		$l_m$ & 0 & 0 & 1 & 0 & 1 & 2\\
		\hline
	\end{tabular}
	\caption{Indexing convention used for the first six PAD basis vectors (indexed by $m$) and their relationships with the $x$ and $y$ derivatives of the PSF (indexed by $k_m$ and $l_m$).}
	\label{tab:indices}	
\end{table}

The PAD basis representation on an even PSF ensures that all basis functions are either even or odd in both $x$ and $y$, i.e., $\phi_m(\vec{x})=(-1)^{p_m}\phi_m(-\vec{x})$, where $p_m=k_m+l_m$ is determined by $m$ according to Table~\ref{tab:indices}. Since this implies $c_{m,n}(\vec{x})=(-1)^{p_{m}+p_n}\big[c_{m,n}(\vec{a}-\vec{x})\big]_{\vec{a}=\vec{\Omega}}$, the density matrix elements of a quantum state of form Eq. \eqref{eq:rhojPAD} from the main text can be written as
\begin{equation}
	d_{j,m,n}=(-1)^{p_m+p_n}\bigg[\frac{1}{\mu^2}m_j\bigg(\frac{\vec{x}}{\mu}\bigg)\conv2D  c_{m,n}(\vec{x})\bigg]_{\vec{x}=\vec{\Omega}},
	\label{eq:rhoconv}
\end{equation} 
where the operator $\conv2D$ represents a 2D convolution. By the convolution theorem along two dimensions, 
\begin{equation}
	\frac{1}{\mu^2}m_j\bigg(\frac{\vec{x}}{\mu}\bigg)\conv2D  c_{m,n}(\vec{x})=\mathcal{F}_{\vec{x}}^{-1}\big\{M_j(\vec{X}_{\rm obj})C_{m,n}(\vec{X})\big\},
	\label{eq:convThm}
\end{equation}
where $\mathcal{F}_{\vec{x}}^{-1}$ represents an inverse 2D Fourier transform over the image-plane Fourier domain coordinates $\vec{X}=\{X,Y\}$, where $\vec{X}_{\rm obj}=\mu\vec{X}$, and where $M_j(\vec{X}_{\rm obj})$ and $C_{m,n}(\vec{X})$ denote the respective 2D characteristic functions of $m_j(\vec{x}_{\rm obj})$ and $c_{m,n}(\vec{x})$. 

The characteristic function of the normalized radiant exitance profile is defined by the object-plane Fourier transform
\begin{equation} 
	\begin{aligned}
		M_j(\vec{X}_{\rm obj})&=\iint_{-\infty}^{\infty}m_j(\vec{x}_{\rm obj})e^{-i(\vec{x}_{\rm obj}\cdot\vec{X}_{\rm obj})}d^2\vec{x}_{\rm obj}  \\
		&=\iint_{-\infty}^{\infty}\tilde{m}_j(\vec{x}_{\rm obj})e^{-i\theta(\vec{x}_{\rm obj}\cdot\vec{X}_{\rm obj})}
		d^2\vec{x}_{\rm obj},
	\end{aligned}
	\label{eq:characteristic1}
\end{equation}
where the second line is found by making the coordinate transformation $\vec{x}_{\rm obj}\to\theta\vec{x}_{\rm obj}$ and applying the definition of $\tilde{m}_j(\vec{x}_{\rm obj})$. Taking a two-dimensional Taylor series expansion about $\vec{X}_{\rm obj}=\vec{\Omega}$,
\begin{equation}
	\begin{aligned}
		M_j(\vec{X}_{\rm obj})=&\sum_{k,l=0}^{\infty}\frac{X_{\rm obj}^kY_{\rm obj}^l}{k!l!}\bigg[\frac{\partial^{k+l}M_j(\vec{X}_{\rm obj})}{\partial X_{\rm obj}^k \partial Y_{\rm obj}^l}\bigg]_{\vec{X}_{\rm obj}=\vec{\Omega}} \\
		=&\sum_{k,l=0}^{\infty}\frac{X_{\rm obj}^kY_{\rm obj}^l}{k!l!}(-i\theta)^{k+l}\\
		&\times\iint_{-\infty}^{\infty}x_{\rm obj}^ky_{\rm obj}^l\tilde{m}_j(\vec{x}_{\rm obj})d^2\vec{x}_{\rm obj}\\
		=&\sum_{k,l=0}^{\infty}\frac{X^kY^l}{k!l!}(-i\mu\theta)^{k+l}m_{j,x^ky^l}.
	\end{aligned}
	\label{eq:characteristic2}
\end{equation}
By the linearity of the Fourier transform, the inverse 2D transform in Eq.~\eqref{eq:convThm} can be evaluated term by term in the Taylor series from Eq.~\eqref{eq:characteristic2}. Setting aside all factors with no $\vec{X}$ dependence, the remaining inverse 2D Fourier transform in each term evaluates at the origin of the image plane to
\begin{widetext}
	\begin{equation}
		\Big[\mathcal{F}_{\vec{x}}^{-1}\big\{X^kY^lC_{m,n}(\vec{X})\big\}\Big]_{\vec{x}=\vec{\Omega}}=(-i)^{k+l}\bigg[\frac{\partial^{k+l}c_{m,n}(\vec{x})}{\partial x^k \partial y^l}\bigg]_{\vec{x}=\vec{\Omega}}
		=\frac{(-i)^{k+l}}{\sigma^{k+l}}\bigg[\frac{\partial^{k+l}\tilde{c}_{m,n}(\vec{x})}{\partial x^k \partial y^l}\bigg]_{\vec{x}=\vec{\Omega}},
		\label{eq:IFT2}
	\end{equation}
\end{widetext}
where the second equality comes from the coordinate transformation $\vec{x}\to\sigma\vec{x}$ and applying the definition $\tilde{c}_{m,n}(\vec{x})=c_{m,n}(\sigma\vec{x})$. These steps taken together result in \begin{equation}
	\begin{aligned}
		d_{j,m,n}=&\sum_{k,l=0}^{\infty}(-1)^{p_m+p_n+k+l}\frac{\gamma^{k+l}}{k!l!}\\
		&\times m_{j,x^ky^l} \bigg[\frac{\partial^{k+l}\tilde{c}_{m,n}(\vec{x})}{\partial x^k \partial y^l}\bigg]_{\vec{x}=\vec{\Omega}},
	\end{aligned}
	\label{eq:rhofull}
\end{equation}
where $p_m=k_m+l_m$. 

Since we seek a lowest order result in the regime where $\gamma\ll 1$, we will focus on the terms in $d_{j,m,n}$ up to $O\big(\gamma^2\big)$, recalling the assumption that the first moments in $x$ and $y$ of the normalized radiant exitance distributions are identical (i.e., $m_{j,x}=0$ and $m_{j,y}=0$, and therefore $m_{j,xy}=0$, without loss of generality). In the PAD basis, where $\vert \phi_0\rangle=\vert\psi_{\vec{x}}\rangle$, we have  $c_{0,0}(\vec{x})=\abs*{\Gamma(\vec{x})}^2$ and $[c_{m,n}(\vec{x})]_{\vec{x}=\vec{\Omega}}=\delta_{m,0}\delta_{n,0}$. We therefore find that a state $\rho_j$ can be expanded in powers of $\gamma$ as
\begin{equation}
	\begin{aligned}
		\rho_j=&\outerproduct{\phi_0}{\phi_0}+\frac{\gamma^2}{2}\sum\limits_{m,n=0}^{\infty}\outerproduct{\phi_{m}}{\phi_{n}}(-1)^{p_m+p_n}\\
		&\times\Bigg(m_{j,x^2}\bigg[\frac{\partial^2 \tilde{c}_{m,n}(\vec{x})}{\partial x^2}\bigg]_{\vec{x}=\vec{\Omega}}\\
		&+m_{j,y^2}\bigg[\frac{\partial^2 \tilde{c}_{m,n}(\vec{x})}{\partial y^2}\bigg]_{\vec{x}=\vec{\Omega}}\Bigg)+O(\gamma^3).
	\end{aligned}
	\label{eq:rhoapprox}
\end{equation}
To analyze the states $\rho_1$ and $\rho_2$, we must evaluate the second-order derivatives in Eq.~\eqref{eq:rhoapprox}. Using Eq.~\eqref{eq:cmntilde}, we find
\begin{equation}
	\begin{aligned}
		\bigg[\frac{\partial^2\tilde{c}_{m,n}(\vec{x})}{\partial x^2}\bigg]_{\vec{x}=\vec{\Omega}}=&\bigg[\frac{\partial^2}{\partial x^2}\Big(\innerproduct{\tilde{\phi}_{m}}{\tilde{\psi}_{\vec{x}}}\innerproduct{\tilde{\psi}_{\vec{x}}}{\tilde{\phi}_{n}}\Big)\bigg]_{\vec{x}=\vec{\Omega}}\\
		=&\innerproduct{\tilde{\phi}_{m}}{\partial_{x^{2}}\tilde{\psi}_{\vec{\Omega}}}\innerproduct{\tilde{\psi}_{\vec{\Omega}}}{\tilde{\phi}_{n}}\\
		&+2\innerproduct{\tilde{\phi}_{m}}{\partial_{x}\tilde{\psi}_{\vec{\Omega}}}\innerproduct{\partial_{x}\tilde{\psi}_{\vec{\Omega}}}{\tilde{\phi}_{n}}\\
		&+\innerproduct{\tilde{\phi}_{m}}{\tilde{\psi}_{\vec{\Omega}}}\innerproduct{\partial_{x^{2}}\tilde{\psi}_{\vec{\Omega}}}{\tilde{\phi}_{n}}
	\end{aligned}
	\label{eq:D2cmn_expanded}
\end{equation}
and the analogous quantity involving derivatives with respect to $y$, where we have defined PSF derivative vectors 
\begin{equation}
	\vert\partial_{x^{k_m}y^{l_m}}\tilde{\psi}_{\vec{\Omega}}\rangle=\bigg[\iint_{-\infty}^{\infty}\frac{\partial^{k_m+l_m}}{\partial x^{k_m}\partial y^{l_m}}\tilde{\psi}(\vec{a}-\vec{x}) \ket{\vec{a}}d^2\vec{a}\bigg]_{\vec{x}=\vec{\Omega}}.
	\label{eq:derivative_kets}
\end{equation}

The inner products in Eq.~\eqref{eq:D2cmn_expanded} can be evaluated by performing the Gram-Schmidt orthogonalization procedure to relate the PAD basis functions to the PSF. The Gram-Schmidt process is equivalent to a QR decomposition $\Psi=\Phi W$ in the single-photon Hilbert space $\mathcal{H}^{(1)}$, where the PSF derivative vectors $	\vert\partial_{x^{k_m}y^{l_m}}\tilde{\psi}_{\vec{\Omega}}\rangle$ define the columns of $\Psi$, $\Phi$ is a unitary matrix with the PAD basis vectors $\vert\tilde{\phi}_m\rangle$ as its columns, and $W$ is a non-singular upper diagonal matrix. In the PAD basis representation, $\Phi$ is an infinite-dimensional identity matrix, so the columns of $\Psi=W$ give the PSF derivative vectors in terms of the PAD basis vectors. This decomposition leads to a natural indexing of the PAD basis elements (Table \ref{tab:indices}) according to ascending total number of derivatives $p_m$ taken in the vectors $\vert\partial_{x^{k_m}y^{l_m}}\tilde{\psi}_{\vec{\Omega}}\rangle$. 

Performing the Gram-Schmidt procedure defines the PAD basis vectors
\begin{equation}
	\vert\tilde{\phi}_{m}\rangle=\sum\limits_{n=0}^{m}w_{n,m} \vert\partial_{x^{k_n}y^{l_n}}\tilde{\psi}_{\vec{\Omega}}\rangle,
	\label{eq:GS-vectors}
\end{equation}
where the coefficients
\begin{equation}
	\begin{aligned}
		w_{0,0}=&1\\
		w_{1,1}=&\frac{1}{\sqrt{\innerproduct{\partial_{x}\tilde{\psi}_{\vec{\Omega}}}{\partial_{x}\tilde{\psi}_{\vec{\Omega}}}}}\\
		w_{2,2}=&\frac{1}{\sqrt{\innerproduct{\partial_{y}\tilde{\psi}_{\vec{\Omega}}}{\partial_{y}\tilde{\psi}_{\vec{\Omega}}}}}\\
		w_{0,3}=&\frac{-\innerproduct{\tilde{\psi}_{\vec{\Omega}}}{\partial_{x^{2}}\tilde{\psi}_{\vec{\Omega}}}}{\sqrt{\innerproduct{\partial_{x^{2}}\tilde{\psi}_{\vec{\Omega}}}{\partial_{x^{2}}\tilde{\psi}_{\vec{\Omega}}}-\abs{\innerproduct{\tilde{\psi}_{\vec{\Omega}}}{\partial_{x^{2}}\tilde{\psi}_{\vec{\Omega}}}}^2}}\\
		w_{3,3}=&\frac{1}{\sqrt{\innerproduct{\partial_{x^{2}}\tilde{\psi}_{\vec{\Omega}}}{\partial_{x^{2}}\tilde{\psi}_{\vec{\Omega}}}-\abs{\innerproduct{\tilde{\psi}_{\vec{\Omega}}}{\partial_{x^{2}}\tilde{\psi}_{\vec{\Omega}}}}^2}}\\
		w_{4,4}=&\frac{1}{\sqrt{\innerproduct{\partial_{xy}\tilde{\psi}_{\vec{\Omega}}}{\partial_{xy}\tilde{\psi}_{\vec{\Omega}}}}}\\
		w_{0,5}=&\frac{\innerproduct{\tilde{\psi}_{\vec{\Omega}}}{\partial_{x^{2}}\tilde{\psi}_{\vec{\Omega}}}\innerproduct{\tilde{\phi}_3}{\partial_{y^{2}}\tilde{\psi}_{\vec{\Omega}}}-\innerproduct{\tilde{\psi}_{\vec{\Omega}}}{\partial_{y^{2}}\tilde{\psi}_{\vec{\Omega}}}}{\sqrt{\innerproduct{\partial_{y^{2}}\tilde{\psi}_{\vec{\Omega}}}{\partial_{y^{2}}\tilde{\psi}_{\vec{\Omega}}}-\abs{\innerproduct{\tilde{\psi}_{\vec{\Omega}}}{\partial_{y^{2}}\tilde{\psi}_{\vec{\Omega}}}}^2-\abs{\innerproduct{\tilde{\phi}_3}{\partial_{y^{2}}\tilde{\psi}_{\vec{\Omega}}}}^2}}\\
		w_{3,5}=&\frac{-\innerproduct{\tilde{\phi}_3}{\partial_{y^{2}}\tilde{\psi}_{\vec{\Omega}}}}{\sqrt{\innerproduct{\partial_{y^{2}}\tilde{\psi}_{\vec{\Omega}}}{\partial_{y^{2}}\tilde{\psi}_{\vec{\Omega}}}-\abs{\innerproduct{\tilde{\psi}_{\vec{\Omega}}}{\partial_{y^{2}}\tilde{\psi}_{\vec{\Omega}}}}^2-\abs{\innerproduct{\tilde{\phi}_3}{\partial_{y^{2}}\tilde{\psi}_{\vec{\Omega}}}}^2}}\\
		w_{5,5}=&\frac{-1}{\sqrt{\innerproduct{\partial_{y^{2}}\tilde{\psi}_{\vec{\Omega}}}{\partial_{y^{2}}\tilde{\psi}_{\vec{\Omega}}}-\abs{\innerproduct{\tilde{\psi}_{\vec{\Omega}}}{\partial_{y^{2}}\tilde{\psi}_{\vec{\Omega}}}}^2-\abs{\innerproduct{\tilde{\phi}_3}{\partial_{y^{2}}\tilde{\psi}_{\vec{\Omega}}}}^2}}
	\end{aligned}
	\label{eq:winvs}
\end{equation}
give the matrix elements of $W^{-1}$ in the PAD basis. All other coefficients $w_{m,n}$ for $0\leq m\leq 5$ and $n\geq m$ are zero due to mismatched even-odd parity between the integer pairs $(k_m,k_n)$ and/or $(l_m,l_n)$ in Eq.~\eqref{eq:GS-vectors}. In matrix form,
\begin{equation}
	W^{-1}=
	\begin{pmatrix}
		w_{0,0} & 0 & 0 & w_{0,3} &0 & w_{0,5} &\cdots\\
		0 & w_{1,1} & 0 & 0 & 0 & 0 &\cdots\\
		0 & 0 & w_{2,2} & 0 & 0 & 0 &\cdots\\
		0 & 0 & 0 & w_{3,3} & 0 & w_{3,5} &\cdots\\
		0 & 0 & 0 & 0 &  w_{4,4} & 0 &\cdots \\
		0 & 0 & 0 & 0 & 0 & w_{5,5} &\cdots \\
		\vdots &\vdots &\vdots  &\vdots & \vdots &\vdots& \ddots
	\end{pmatrix},
	\label{eq:Winv}
\end{equation}
where we display just the subspace $\mathcal{H}_6^{(1)}$ corresponding to the first six PAD basis vectors according to Table~\ref{tab:indices}. Since the Schur complement of any block of an upper triangular matrix is simply given by that matrix block, the inverse of any diagonal block of an upper triangular matrix is equal to the same diagonal block of the inverse of the matrix. Inverting $W^{-1}$, we find that the matrix  

\begin{equation}
	W=
	\begin{pmatrix}
		W_{0,0} & 0 & 0 & W_{0,3} &0 & W_{0,5} &\cdots\\
		0 & W_{1,1} & 0 & 0 & 0 & 0 &\cdots\\
		0 & 0 & W_{2,2} & 0 & 0 & 0 &\cdots\\
		0 & 0 & 0 & W_{3,3} & 0 & W_{3,5} &\cdots\\
		0 & 0 & 0 & 0 &  W_{4,4} & 0 &\cdots \\
		0 & 0 & 0 & 0 & 0 & W_{5,5} &\cdots \\
		\vdots &\vdots &\vdots  &\vdots & \vdots &\vdots& \ddots
	\end{pmatrix}
	\label{eq:W}
\end{equation}
with
\begin{equation}
	\begin{aligned}
		W_{0,0}=&\frac{1}{w_{0,0}}=1\\
		W_{1,1}=&\frac{1}{w_{1,1}}=\sqrt{\innerproduct{\partial_{x}\tilde{\psi}_{\vec{\Omega}}}{\partial_{x}\tilde{\psi}_{\vec{\Omega}}}}\\
		W_{2,2}=&\frac{1}{w_{2,2}}=\sqrt{\innerproduct{\partial_{y}\tilde{\psi}_{\vec{\Omega}}}{\partial_{y}\tilde{\psi}_{\vec{\Omega}}}}\\
		W_{0,3}=&-\frac{w_{0,3}}{w_{0,0}w_{3,3}}=\innerproduct{\tilde{\psi}_{\vec{\Omega}}}{\partial_{x^{2}}\tilde{\psi}_{\vec{\Omega}}}\\
		W_{3,3}=&\frac{1}{w_{3,3}}=\sqrt{\innerproduct{\partial_{x^{2}}\tilde{\psi}_{\vec{\Omega}}}{\partial_{x^{2}}\tilde{\psi}_{\vec{\Omega}}}-\abs{\innerproduct{\tilde{\psi}_{\vec{\Omega}}}{\partial_{x^{2}}\tilde{\psi}_{\vec{\Omega}}}}^2}\\
		W_{4,4}=&\frac{1}{w_{4,4}}=\sqrt{\innerproduct{\partial_{xy}\tilde{\psi}_{\vec{\Omega}}}{\partial_{xy}\tilde{\psi}_{\vec{\Omega}}}}\\
		W_{0,5}=&\frac{w_{0,3}w_{3,5}-w_{0,5}w_{3,3}}{w_{0,0}w_{3,3}w_{5,5}}=\innerproduct{\tilde{\psi}_{\vec{\Omega}}}{\partial_{y^{2}}\tilde{\psi}_{\vec{\Omega}}}\\
		W_{3,5}=&-\frac{w_{3,5}}{w_{3,3}w_{5,5}}\\
		=&\innerproduct{\partial_{x^{2}}\tilde{\psi}_{\vec{\Omega}}}{\partial_{y^{2}}\tilde{\psi}_{\vec{\Omega}}}-\innerproduct{\partial_{x^{2}}\tilde{\psi}_{\vec{\Omega}}}{\tilde{\psi}_{\vec{\Omega}}}\innerproduct{\tilde{\psi}_{\vec{\Omega}}}{\partial_{y^{2}}\tilde{\psi}_{\vec{\Omega}}}\\
		W_{5,5}=&\frac{1}{w_{5,5}}\\
		=&\sqrt{\innerproduct{\partial_{y^{2}}\tilde{\psi}_{\vec{\Omega}}}{\partial_{y^{2}}\tilde{\psi}_{\vec{\Omega}}}-\abs{\innerproduct{\tilde{\psi}_{\vec{\Omega}}}{\partial_{y^{2}}\tilde{\psi}_{\vec{\Omega}}}}^2-\abs{\innerproduct{\tilde{\phi}_3}{\partial_{y^{2}}\tilde{\psi}_{\vec{\Omega}}}}^2}
	\end{aligned}
	\label{eq:ws}
\end{equation}
represents on its columns the non-orthonormal PSF derivative vectors in the orthonormal PAD basis, as in
\begin{equation}
	\vert\partial_{x^{k_m}y^{l_m}}\tilde{\psi}_{\vec{\Omega}}\rangle=\sum\limits_{n=0}^{m}W_{n,m}\vert\tilde{\phi}_{n}\rangle.
	\label{eq:derivative-vectors}
\end{equation}
These steps show that the $m^{\rm th}$ PSF derivative vector depends on only the basis vectors $\vert\tilde{\phi}_n\rangle$ for which $n\leq m$. 

Using Eq.~\eqref{eq:derivative-vectors} and $\langle \tilde{\phi}_m\vert\tilde{\phi}_n\rangle=\delta_{m,n}$, we rewrite the density matrix elements [Eq.~\eqref{eq:rhofull}] in the PAD basis as
\begin{equation}
	\begin{aligned}
		d_{j,0,0}=&1+\textrm{Re}\big(m_{j,x^2}W_{0,3}+m_{j,y^2}W_{0,5}\big)\gamma^2+O(\gamma^3)\\
		=&1-(m_{j,x^2}\Gamma_{x^2}+m_{j,y^2}\Gamma_{y^2})\gamma^2+O(\gamma^3)\\
		d_{j,1,1}=&m_{j,x^2}W_{1,1}^2\gamma^2+O(\gamma^3)\\
		=&m_{j,x^2}\Gamma_{x^2}\gamma^2+O(\gamma^3)\\
		d_{j,2,2}=&m_{j,y^2}W_{2,2}^2\gamma^2+O(\gamma^3)\\
		=&m_{j,y^2}\Gamma_{y^2}\gamma^2+O(\gamma^3)\\
		d_{j,3,0}=&\frac{1}{2}\Big(m_{j,x^2}W_{3,3}+m_{j,y^2}W_{3,5}\Big)\gamma^2+O(\gamma^3)\\
		d_{j,0,3}=&\frac{1}{2}\Big(m_{j,x^2}W^\ast_{3,3}+m_{j,y^2}W^\ast_{3,5}\Big)\gamma^2+O(\gamma^3)\\
		d_{j,5,0}=&\frac{1}{2}m_{j,y^2}W_{5,5}\gamma^2+O(\gamma^3)\\
		d_{j,0,5}=&\frac{1}{2}m_{j,y^2}W^\ast_{5,5}\gamma^2+O(\gamma^3),
	\end{aligned}
	\label{eq:rho_j_sparse}
\end{equation}
where all other matrix elements are $O\big(\gamma^3\big)$. The relationships $\textrm{Re}(W_{0,3})=-\Gamma_{x^2}$ and $\textrm{Re}(W_{0,5})=-\Gamma_{y^2}$ are given as an identity in Ref. \cite{Kerviche2017a}, while $W_{1,1}^2=\Gamma_{x^2}$ (and likewise $W_{2,2}^2=\Gamma_{y^2}$) can be derived by taking 2nd derivatives on either side of the PSF normalization condition 
and exchanging the order of integration and differentiation:
\begin{equation}
	\begin{split}
		0
		=&\iint_{-\infty}^{\infty}\frac{\partial^2}{\partial x^2}\abs*{\tilde{\psi}(\vec{x})}^2d^2\vec{x}\\
		=&\iint_{-\infty}^{\infty}2\abs*{\frac{\partial\tilde{\psi}(\vec{x})}{\partial x}}^2+2\textrm{Re}\bigg(\tilde{\psi}^\ast(\vec{x})\frac{\partial^2\tilde{\psi}(\vec{x})}{\partial x^2}\bigg)d^2\vec{x}\\
		=&2(W_{1,1}^2-\Gamma_{x^2}).
	\end{split}
\end{equation} 
Note that if the PSF is separable in $x$ and $y$ i.e., $\psi(\vec{x})=\Xi(x)\Upsilon(y)$, then $W_{3,5}=0$.

With the states $\rho_1$ and $\rho_2$ [Eq.~\eqref{eq:rhojPAD} in the main text] represented in the finite dimensional PAD basis defined on the truncated subspace $\mathcal{H}_6^{(1)}$, we can directly evaluate the quantum Chernoff exponent [Eq.~\eqref{eq:QuantumChernoff} in the main text] to lowest non-vanishing order in $\gamma$ using our accompanying work on perturbation theory for common entropic and distance measures in quantum information theory \cite{Grace2021}. We first decompose both density matrices into $\rho_j=\rho_0+\nu_j$, where $\rho_0=\outerproduct{\phi_0}{\phi_0}$ is a pure state and $\nu_1$ and $\nu_2$ are zero-trace perturbation matrices with spectral norms $\norm{\nu_j}=O(\gamma^2)$. We additionally block decompose the truncated Hilbert space as $\mathcal{H}_6^{(1)}=\mathcal{H}_1^{(1)}\oplus\mathcal{H}_5^{(1)}$, where $\mathcal{H}_1^{(1)}$ is the 1-dimensional subspace of $\mathcal{H}_6^{(1)}$ corresponding to the support of the pure state $\rho_0$ and $\mathcal{H}_5^{(1)}$ is the kernel of $\rho_0$ on $\mathcal{H}_6^{(1)}$. We denote the resulting matrix decompositions using the block matrices
\begin{equation}
	\begin{split}
		\rho_0 =& 
		\begin{pmatrix}
			1 & 0\\
			0 & 0
		\end{pmatrix}\\
		\nu_j=&
		\begin{pmatrix}
			\nu_{j,\rm B} & \nu_{j,\rm C}\\
			\nu_{j,\rm C}^{\dagger} & \nu_{j,\rm D}
		\end{pmatrix}.
	\end{split}
	\label{eq:BlockDecomposition_nu}
\end{equation}
Following the perturbation theory for perturbations that extend the support of the original states \cite{Grace2021}, the QCE
\begin{equation}
	\begin{split}
		\xi_{\rm Q}^{(1)}=& -\min_{s\in[0,1]}s\Tr[\nu_{1,\rm B}]+(1-s)\Tr[\nu_{2, \rm B}]\\
		&+\Tr\big[\nu_{1,\rm D}^s\nu_{2, \rm D}^{1-s}\big]+O\big(\max(\norm{\nu_1},\norm{\nu_2})^2\big)
	\end{split}
	\label{eq:QCEperturbation}
\end{equation}
can be easily evaluated in the PAD basis by reading off the matrix elements of $\nu_{1,\rm B}$, $\nu_{2, \rm B}$, $\nu_{1,\rm D}$ and $\nu_{2, \rm D}$ using Eq.~\eqref{eq:rho_j_sparse} because $\nu_{1,\rm D}$ and $\nu_{2, \rm D}$ are diagonal. The result is Eq.~\eqref{eq:GeneralQCE} in the main text.

\subsection{Direct Imaging for Binary Object Discrimination}
For a continuously valued measurement (i.e., where the outcome space $\mathcal{Z}$ is continuous), the classical Chernoff exponent takes the form 
\begin{equation}
	\xi_{\textrm{Direct}}^{(1)}=-\log\Big[\min\limits_{0\leq s \leq 1} Q_s\Big]
	\label{eq:ContinuousChernoff}
\end{equation}
with
\begin{equation}
	Q_s=\int_{\mathcal{Z}}P(z\vert\rho_1)^s P(z\vert\rho_2)^{1-s}dz.
	\label{eq:ContinuousQs}
\end{equation}
Ideal direct imaging implements a POVM $\{\Pi_{\vec{x}}\}_{\mathbb{R}^2}$, with elements $\Pi_{\vec{x}}=\outerproduct{\vec{x}}{\vec{x}}$ corresponding to photon arrival locations $\vec{x}$ in the outcome space $\mathbb{R}^2$. The classical Chernoff exponent for direct imaging can therefore be found using
\begin{equation}
	Q_s=\iint_{-\infty}^{\infty}\big(\sigma^2 P(\sigma\vec{x}\vert\rho_1)\big)^s\big(\sigma^2 P(\sigma\vec{x}\vert\rho_2)\big)^{1-s}d^2\vec{x},
	\label{eq:QsDirect}
\end{equation}
where we made the coordinate transformation $\vec{x}\to\sigma\vec{x}$ and distributed the factor $\sigma^2$.

From Eq.~\eqref{eq:rhoj}, we have
\begin{equation}
	\begin{aligned}
		P(\vec{x}\vert\rho_j)=&\Tr\big[\rho_j \outerproduct{\vec{x}}{\vec{x}}d^2\vec{x}\big] \\
		=&\iint_{-\infty}^\infty \frac{1}{\mu^2}m_j\bigg(\frac{\vec{a}}{\mu}\bigg)\langle \vec{x}\vert\psi_{\vec{a}}\rangle\langle\psi_{\vec{a}}\vert\vec{x}\rangle d^2\vec{a}\\
		=&\frac{1}{\mu^2}m_j\bigg(\frac{\vec{x}}{\mu}\bigg)\ast\ast\abs*{\psi(\vec{x})}^2,
	\end{aligned}
	\label{eq:DirectP}
\end{equation} 
which is consistent with the standard result from classical image science that the image plane intensity distribution in an incoherent imaging context is given by a 2D convolution of the object radiant exitance and the incoherent PSF \cite{Goodman2005}. By the convolution theorem, $(1/\mu^2)~m_j(\vec{x}/\mu)\conv2D \abs*{\psi(\vec{x})}^2=\mathcal{F}_{\vec{x}}^{-1}\big\{M_j(\vec{X}_{\rm obj})\Psi(\vec{X})\big\}$, where $\Psi(\vec{X})$ is the 2D characteristic function of the incoherent PSF $\abs*{\psi(\vec{x})}^2$. Using Eq.~\eqref{eq:characteristic2} and evaluating the inverse 2D Fourier transform via similar methods to those used for Eq.~\eqref{eq:IFT2}, we find that the relevant probability density functions for the integrand of Eq.~\eqref{eq:QsDirect} have the form
\begin{equation}
	\begin{aligned}
		P(\sigma\vec{x}\vert\rho_j)=&\frac{1}{\sigma^2}\sum_{k,l=0}^{\infty}(-1)^{k+l}\frac{\gamma^{k+l}}{k!l!}m_{j,x^ky^l}\psi_{x^ky^l}(\vec{x})
	\end{aligned}
	\label{eq:DirectPSeries}
\end{equation}
where the derivatives of the incoherent PSF
\begin{equation}
	\psi_{x^ky^l}(\vec{x})=\frac{\partial^{k+l}\abs{\tilde{\psi}(\vec{x})}^2}{\partial x^k \partial y^l}
\end{equation} 
were defined in the main text.

In order to calculate the Chernoff exponent in the limit $\gamma\ll 1$, we generalize the methods from Ref. \cite{Lu2018} and expand the integrand of $Q_s$ using the Taylor series
\begin{equation}
	\begin{aligned}
		f(\gamma)^sg(\gamma)^{1-s}=&f_0+(sf_2+(1-s)g_2)\gamma^2\\
		&+(sf_3+(1-s)g_3)\gamma^3\\
		&+\bigg[sf_4+(1-s)g_4\\
		&-\frac{1}{2f_0}s(1-s)(f_2-g_2)^2\bigg]\gamma^4 + O(\gamma^5)
	\end{aligned}
	\label{eq:IntegrandSeries}
\end{equation}
where $f(\gamma)\equiv\sum_{n=0}^\infty f_n\gamma^n=\sigma^2 P(\sigma\vec{x}\vert\rho_0)$ and $g(\gamma)\equiv\sum_{n=0}^\infty g_n\gamma^n=\sigma^2 P(\sigma\vec{x}\vert\rho_1)$, and where we used $g_0=f_0$ and $g_1=f_1=0$ since $m_{j,x}=m_{j,y}=m_{j,xy}=0$. First, we consider only terms in $Q_s$ up to third order in $\gamma$. Using Eq. \eqref{eq:DirectPSeries} to determine the coefficients $f_n$ and $g_n$, Eqs.~\eqref{eq:QsDirect} and \eqref{eq:IntegrandSeries} result in
\begin{equation}
	\begin{aligned}
		Q_s=&\iint_{-\infty}^{\infty}\abs{\tilde{\psi}(\vec{x})}^2\\
		&+\frac{1}{2}\big[\big(sm_{1,x^2}+(1-s)m_{2,x^2}\big)\psi_{x^2}(\vec{x})\\
		&+\big(sm_{1,y^2}+(1-s)m_{2,y^2}\big)\psi_{y^2}(\vec{x})\big]\gamma^2\\
		&+\frac{1}{6}\big[\big(sm_{1,x^3}+(1-s)m_{2,x^3}\big)\psi_{x^3}(\vec{x})\\
		&+\big(sm_{1,y^3}+(1-s)m_{2,y^3}\big)\psi_{y^3}(\vec{x})\big]\gamma^3d^2\vec{x}+O(\gamma^4).
	\end{aligned}
	\label{eq:ChernoffDirect3}
\end{equation}
After distributing the integral to each of the terms, we note that
\begin{equation}
	\iint_{-\infty}^{\infty}\psi_{x^ky^l}(\vec{x})d^2\vec{x}=\delta_{k,0}\delta_{l,0},
	\label{eq:PSFidentity}
\end{equation} 
which can be shown by exchanging the order of integration and differentiation and using the normalization condition on $\tilde{\psi}(\vec{x})$ \cite{Lu2018}. Thus, $Q_s=1+O(\gamma^4)$ for all values of $s$, and $\xi_{\textrm{Direct}}^{(1)}=O(\gamma^4)$ in the sub-Rayleigh limit, while the quantum Chernoff exponent has nonzero $\gamma^2$ terms [Eq. \eqref{eq:GeneralQCE} in the main text]. This implies at least a quadratic scaling difference between the achievable error exponent when using direct imaging compared with the quantum limit for any sub-Rayleigh binary object discrimination task with an arbitrary 2D aperture.

To get a closed form expression for the direct detection classical Chernoff exponent to lowest non-zero order in $\gamma\ll 1$ for a restricted class of PSFs (see below), we expand $Q_s$ to fourth order in $\gamma$ using all of the terms given in Eq.~\eqref{eq:IntegrandSeries}. After integrating the $\gamma^2$ and $\gamma^3$ terms of $Q_s$ to zero as before, we have
\begin{equation}
	\begin{aligned}
		Q_s=&\iint_{-\infty}^{\infty}\abs{\tilde{\psi}(\vec{x})}^2\\
		&+\frac{1}{24}\bigg[\big(sm_{1,x^4}+(1-s)m_{2,x^4}\big)\psi_{x^4}(\vec{x})\\
		&+6\big(sm_{1,x^2y^2}+(1-s)m_{2,x^2y^2}\big)\psi_{x^2y^2}(\vec{x})\\
		&+\big(sm_{1,y^4}+(1-s)m_{2,y^4}\big)\psi_{y^4}(\vec{x})\\
		&-\frac{3s(1-s)}{\abs{\tilde{\psi}(\vec{x})}^2}\Big((m_{1,x^2}-m_{2,x^2})\psi_{x^2}(\vec{x})\\
		&+(m_{1,y^2}-m_{2,y^2})\psi_{y^2}(\vec{x})\Big)^2\bigg]\gamma^4d^2\vec{x}+O(\gamma^5).
	\end{aligned}
	\label{eq:ChernoffDirect4}
\end{equation}
Applying Eq.~\eqref{eq:PSFidentity}, we find $Q_s=1-s(1-s)\mathcal{K}\gamma^4/8+O(\gamma^5)$, where $\mathcal{K}$ is given by 
\begin{equation}
	\begin{split}
		\mathcal{K}=&\iint_{-\infty}^{\infty}\abs{\tilde{\psi}(\vec{x})}^{-2}\Big((m_{1,x^2}-m_{2,x^2})\psi_{x^2}(\vec{x})\\
		&+(m_{1,y^2}-m_{2,y^2})\psi_{y^2}(\vec{x})\Big)^2d^2\vec{x}.
	\end{split}
	\label{eq:K}
\end{equation}
Minimizing $Q_s$ is then equivalent to maximizing $s(1-s)$ over $0\leq s\leq 1$, so $s=1/2$ and the Chernoff exponent is given by
\begin{equation}
	\xi_{\textrm{Direct}}^{(1)}=\frac{\mathcal{K}}{32}\gamma^4+O(\gamma^5).
	\label{eq:CEDirectGeneral}
\end{equation}
If the PSF is separable in $x$ and $y$, i.e., $\psi(\vec{x})=\Xi(x)\Upsilon(y)$, we have
\begin{equation}
	\begin{split}
		\psi_{x^2}(\vec{x})\psi_{y^2}(\vec{x})d^2\vec{x}=&\abs{\tilde{\Upsilon}(y)}^2\frac{d\abs{\tilde{\Xi}(x)}^2}{dx^2}\abs{\tilde{\Xi}(x)}^2\frac{d\abs{\tilde{\Upsilon}(y)}^2}{dx^2}\\
		=&\abs{\tilde{\psi}(\vec{x})}^2\psi_{x^2y^2}(\vec{x}),
	\end{split}
	\label{eq:SeparablePSFIdentity}
\end{equation}
and $\iint_{-\infty}^{\infty}\abs{\tilde{\psi}(\vec{x})}^{-2}\psi_{x^2}(\vec{x})\psi_{y^2}(\vec{x})d^2\vec{x}=0$ follows from Eq.~\eqref{eq:PSFidentity}. Therefore, Eq.~\eqref{eq:CEDirectGeneralSeparable} in the main text can be found by expanding terms inside the integral of Eq.~\eqref{eq:K}. 

A very important caveat is that this closed form expression for the direct imaging Chernoff exponent [Eq.~\eqref{eq:CEDirectGeneral} or Eq.~\eqref{eq:CEDirectGeneralSeparable} in the main text] does not hold for all PSFs. Namely, the presence of the factor $\abs{\tilde{\psi}(\vec{x})}^{-2}$ in the integrand of $\mathcal{K}$ (or $\mathcal{K}_x$ and $\mathcal{K}_y$) results in a diverging integral if the coherent PSF $\psi(\vec{x})$ is exactly zero at any point $\vec{x}\in\mathbb{R}^2$. Most significantly, this includes the Airy disk PSF associated with a hard circular aperture \cite{Goodman2005}, a ubiquitous aperture geometry for realistic imaging systems. A similar situation was encountered in Ref.~\cite{Paur2019}, which pointed out that the sub-Rayleigh scaling of the Fisher information for estimating an object parameter cannot be evaluated by expanding the integrand as a Taylor series and integrating individual terms when there are zeros in the PSF. Likewise, our integration of the individual terms of Eq.~\eqref{eq:ChernoffDirect4} is invalid for such PSFs. However, our result differs in kind from that of Ref.~\cite{Paur2019}, which reported a factor of $\gamma$ (in our notation) improvement in the scaling of the direct imaging Fisher information using a hard aperture compared with that using a Gaussian attenuated aperture model. In the case of binary hypothesis testing, we find that the integrals of the individual terms of $Q_s$ up to third order in $\gamma$ always converge [Eq. \eqref{eq:ChernoffDirect3}], so the result $\xi_{\textrm{Direct}}^{(1)}=O(\gamma^4)$ holds for any PSF. Therefore, while we cannot write down a closed form expression for $\xi_{\textrm{Direct}}^{(1)}$ when using an aperture with zeros in its PSF, the scaling gap between the classical and quantum Chernoff exponents is always at least quadratic in the sub-Rayleigh limit, and PSFs with zeros give no fundamental advantage for sub-Rayleigh object discrimination. 

\subsection{Mode-Sorting Measurements for Binary Object Discrimination}
We first consider a binary projective measurement with POVM elements $\Pi_0=\outerproduct{\psi_{\vec{\Omega}}}{\psi_{\vec{\Omega}}}$ and $\Pi_1= \mathcal{I}-\outerproduct{\psi_{\vec{\Omega}}}{\psi_{\vec{\Omega}}}$, which can be implemented by a 2D BSPADE device \cite{Ang2017}. From Eqs.~\eqref{eq:rhojPAD} in the main text and Eq.~\eqref{eq:rho_j_sparse},
\begin{equation}
	\begin{aligned}
		P(0\vert\rho_j)=&d_{j,0,0}\\
		=& 1-(m_{j,x^2}\Gamma_{x^2}+m_{j,y^2}\Gamma_{y^2})\gamma^2+O(\gamma^3)\\
		P(1\vert\rho_j)=&1-d_{j,0,0}\\
		=& (m_{j,x^2}\Gamma_{x^2}+m_{j,y^2}\Gamma_{y^2})\gamma^2+O(\gamma^3).
	\end{aligned}
	\label{eq:BSPADEP}
\end{equation}
We apply Eq.~\eqref{eq:Chernoff} from the main text and utilize the first order expansion $(1+u)^v=1+vu+O(u^2)$ to simplify the terms $P(0\vert\rho_1)^s$ and $P(0\vert\rho_2)^{1-s}$. Expanding the logarithm in that equation using $\ln(1+u)=u+O(u^2)$, the classical Chernoff exponent for the 2D BSPADE measurement is given by
\begin{equation}
	\begin{aligned}
		\xi_{\rm BSPADE}^{(1)}=&\max\limits_{0\leq s \leq 1}\Big[s(m_{1,x^2}\Gamma_{x^2}+m_{1,y^2}\Gamma_{y^2})\\
		&+(1-s)(m_{2,x^2}\Gamma_{x^2}+m_{2,y^2}\Gamma_{y^2})\\
		&-(m_{1,x_{\rm obj}^2}\Gamma_{x^2}+m_{1,y_{\rm obj}^2}\Gamma_{y^2})^s\\
		&\times(m_{2,x^2}\Gamma_{x^2}+m_{2,y^2}\Gamma_{y^2})^{1-s}\Big]\gamma^2\\
		&+O(\gamma^3).
	\end{aligned}
	\label{eq:BSPADECE}
\end{equation}
Unlike hypothesis tests between a point source and an arbitrary object, the 2D BSPADE measurement does not in general achieve the quantum limit [Eq. \eqref{eq:GeneralQCE} in the main text] in the sub-Rayleigh limit for a binary hypothesis test between any two arbitrary objects. A few steps of algebra show that $\xi_{\rm BSPADE}^{(1)}=\xi_{\rm Q}^{(1)}$ when the two objects have the same ellipticity, i.e., when $m_{1,x^2}/m_{1,y^2}=m_{2,x^2}/m_{2,y^2}$, but when this condition is not met the 2D BSPADE measurement is not quantum-optimal. 

Alternatively, the projectors $\Pi_0=\outerproduct{\phi_{0}}{\phi_{0}}$, $\Pi_1=\outerproduct{\phi_{1}}{\phi_{1}}$, $\Pi_2=\outerproduct{\phi_{2}}{\phi_{2}}$, and $\Pi_3=\mathcal{I}-\Pi_0-\Pi_1-\Pi_2$ form a POVM and are constructed from the first three eigenvectors of the PAD basis and the orthogonal complement on $\mathcal{H}^{(1)}$. From Eq.~\eqref{eq:rho_j_sparse}, the measurement outcome probabilities are
\begin{equation}
	\begin{aligned}
		P(0\vert\rho_j)=&d_{j,0,0}\\
		=& 1-(m_{j,x^2}\Gamma_{x^2}+m_{j,y^2}\Gamma_{y^2})\gamma^2+O(\gamma^3)\\
		P(1\vert\rho_j)=&d_{j,1,1}= m_{j,x^2}\Gamma_{x^2}\gamma^2+O(\gamma^3)\\ 
		P(2\vert\rho_j)=&d_{j,2,2}= m_{j,y^2}\Gamma_{y^2}\gamma^2+O(\gamma^3)\\
		P(3\vert\rho_j)=&1-d_{j,0,0}-d_{j,1,1}-d_{j,2,2}=O(\gamma^3).
	\end{aligned}
	\label{eq:3-SPADEP}
\end{equation}
Since $\Pi_3$ has negligible outcome probability in the limit $\gamma\ll1$, the fourth POVM element can be ignored in a practical implementation designed for sub-Rayleigh imaging. Using Eq.~\eqref{eq:Chernoff} from the main text and the same expansions as before, the CE of the resulting TriSPADE measurement formed by the projectors $\Pi_0$, $\Pi_1$, and $\Pi_2$,
\begin{equation}
	\begin{aligned}
		\xi_{\rm TriSPADE}^{(1)}=&\max\limits_{0\leq s \leq 1}\Big[s(m_{1,x^2}\Gamma_{x^2}+m_{1,y^2}\Gamma_{y^2})\\
		&+(1-s)(m_{2,x^2}\Gamma_{x^2}+m_{2,y^2}\Gamma_{y^2})\\
		&-m_{1,x^2}^sm_{2,x^2}^{1-s}\Gamma_{x^2}\\
		&-m_{1,y^2}^sm_{2,y^2}^{1-s}\Gamma_{y^2}\Big]\gamma^2+O(\gamma^3),
	\end{aligned}
	\label{eq:3-SPADECE}
\end{equation}
is exactly equal to the QCE [Eq.~\eqref{eq:GeneralQCE} in the main text]. 

As an aside, since the Cartesian derivatives of a function that is even in $x$ and $y$ will all be either odd or even, the PAD-basis vectors inherit even or odd parity. The projectors $\Pi_0$, $\Pi_1$, $\Pi_2$ and $\Pi_3$ can therefore be thought of as sorting the 2D even/odd parity of the captured optical field to lowest order in $\gamma$. As a result, the same quantitative performance will be achieved with a 2D inversion-interferometric measurement (SLIVER)~\cite{Nair2016b,Ang2017}. This device may be easier to implement for some applications.

\subsection{Special Cases for Binary Object Discrimination}
Our results must be modified in the special case of two candidate objects that have exactly the same second moment in $x_{\rm obj}$ and $y_{\rm obj}$. In this case, the quantities given in Eqs. \eqref{eq:GeneralQCE}, \eqref{eq:CEDirectGeneral}, \eqref{eq:BSPADECE}, and \eqref{eq:3-SPADECE} are all zero to the orders in $\gamma$ specified, and higher order terms are needed to represent these Chernoff exponents. Let $\kappa$ be the lowest-order moment in $x_{\rm obj}$ and/or $y_{\rm obj}$ that differs between $m_1(\vec{x}_{\rm obj})$ and $m_2(\vec{x}_{\rm obj})$, i.e., $m_{1,x^ky^l}=m_{2,x^ky^l}$ for all $k+l<\kappa$. For the quantum Chernoff exponent, we reverse the order of summation in Eq.~\eqref{eq:rhojPAD} from the main text and in Eq.~\eqref{eq:rhofull} to write the state of an arbitrary 2D incoherent object as
\begin{equation}
	\begin{aligned}
		\rho_j=&\sum_{k,l=0}^{\infty}\frac{\gamma^{k+l}}{k!l!}\sum_{m,n=0}^{\infty}\outerproduct{\phi_{m}}{\phi_{n}}(-1)^{p_m+p_n+k+l}\\
		&\times m_{j,x^ky^l} \bigg[\frac{\partial^{k+l}\tilde{c}_{m,n}(\vec{x})}{\partial x^k \partial y^l}\bigg]_{\vec{x}=\vec{\Omega}}.
	\end{aligned}
	\label{eq:rhoapprox_kappa}
\end{equation}
We then perform the decomposition of Eq.~\eqref{eq:BlockDecomposition_nu}: $\rho_0$ contains all terms with $k+l<\kappa$ in the outer sum of Eq.~\eqref{eq:rhoapprox_kappa}, while $\nu_j=O(\gamma^\kappa)$ contains the terms with $k+l\geq\kappa$. It is clear from the perturbation theory result of Eq.~\eqref{eq:QCEperturbation} that the QCE will be $\xi_{\rm Q}^{(1)}=O\big(\gamma^\kappa\big)$. On the other hand, the CE for direct imaging can be analyzed using a generalization of the series expansion of Eq. \eqref{eq:IntegrandSeries}. If $f(\gamma)=\sum_{n=0}^{\infty}f_n \gamma^n$ and $g(\gamma)=\sum_{n=0}^{\kappa-1}f_n \gamma^n+\sum_{n=\kappa}^{\infty}g_n \gamma^n$, then 
\begin{equation}
	\begin{aligned}
		f(\gamma)^{s}g(\gamma)^{1-s}=&\sum_{n=0}^{\kappa-1}f_n\gamma^n+\sum_{n=\kappa}^{2\kappa-1}(s f_n+(1-s)g_n)\gamma^n\\
		&-\frac{1}{2f_0}s(1-s)(f_{\kappa}-g_{\kappa})^2\gamma^{2\kappa}+O(\gamma^{2\kappa+1}).
	\end{aligned}
	\label{eq:IntegrandSeriesGeneral}
\end{equation}
Applying this series to the integral given by Eqs.~\eqref{eq:QsDirect} and~\eqref{eq:DirectPSeries}, we perform term-by-term integration of $Q_s$ up to order $2\kappa-1$ in $\gamma$ using the identity in Eq.~\eqref{eq:PSFidentity}, resulting in $Q_s=1+O\big(\gamma^{2\kappa}\big)$ and $\xi_{\rm Direct}^{(1)}=O\big(\gamma^{2\kappa}\big)$. We thus find the scaling gap between the QCE and the CE for direct imaging to be of the order $O(\gamma^\kappa)$. The TriSPADE measurement will not be quantum-optimal in this case, and finding an optimal measurement remains an open question for future work.

In the opposite scenario, when the two objects have different first moments in $x_{\rm obj}$ or $y_{\rm obj}$ (i.e., different centroids), we encounter different behavior: the scaling gap between the quantum limit and direct imaging vanishes. For the quantum Chernoff exponent, the states $\rho_j$ can be approximated using Eq.~\eqref{eq:rhoapprox_kappa}, with $\kappa=1$. Since lateral shifts of the reference frame should not affect the quantum Chernoff exponent, it is valid to move the origin of the object plane coordinate system so that the Cartesian first moments of one of the two objects are zero, i.e., $m_{1,x}=m_{1,y}=0$. 
As a result, $\rho_1=\outerproduct{\phi_1}{\phi_1}+O(\gamma^2)$ is a pure state up to first order in $\gamma$, and the QCE is given by $\xi_{\textrm{Q}}^{(1)}=-\log\big[ F(\outerproduct{\phi_1}{\phi_1},\rho_2)\big]+O(\gamma^2)$. Using $\log(1+u)= u+O(u^2)$, the QCE becomes
\begin{equation}
	\begin{split}
		\xi_{\rm Q}^{(1)}=& \Bigg(m_{2,x}\bigg[\frac{\partial\tilde{\Gamma}(\vec{x})}{\partial x}\bigg]_{\vec{x}=\vec{\Omega}}+m_{2,y}\bigg[\frac{\partial\tilde{\Gamma}(\vec{x})}{\partial y}\bigg]_{\vec{x}=\vec{\Omega}}\Bigg)\gamma\\
		&+O(\gamma^2),
	\end{split}
	\label{eq:QCE_differentCentroids}
\end{equation}
which is zero because $\tilde{\Gamma}(\vec{x})$ is even in $x$ and $y$ for a circularly symmetric PSF. Therefore, the lowest nonzero order term of the QCE is still of order $O(\gamma^2)$. On the other hand, for the direct imaging Chernoff exponent, it is trivial to use Eq.~\eqref{eq:IntegrandSeriesGeneral} with $\kappa=1$ to find that $\xi_{\textrm{Direct}}^{(1)}=O(\gamma^2)$. We therefore find that the scaling gap with respect to $\gamma$ between the QCE and the direct imaging CE disappears when the two objects have different centroids. 

\subsection{$M$-ary Object Discrimination}
The $M$-ary QCE and the measurement-specific $M$-CE extend naturally from $M=2$ to $M>2$ candidates via the minimizations $\xi_{\textrm{Q},M}^{(1)}=\min_{i\neq j}\xi_{\textrm{Q},i,j}^{(1)}$ and $\xi_{\textrm{Meas},M}^{(1)}=\min_{i\neq j}\xi_{\textrm{Meas},i,j}^{(1)}$ over all pairwise QCEs and CEs, respectively \cite{Li2016}. We analyze the $M$-ary QCE by approximating the pairwise QCEs using the quantum Bhattacharyya bound \cite{Pirandola2008}:
\begin{equation}
	\xi_{\textrm{Q},i,j}^{(1)}\geq\xi_{\textrm{B},i,j}^{(1)}=-\log\Big[\Tr\big(\rho_i^{1/2}\rho_j^{1/2}\big)\Big].
	\label{eq:QBB}
\end{equation}
The quantum Bhattacharyya bound is proven to be tight, matching the quantum Chernoff bound, for any two quantum states that share the same Hilbert space support and differ by a vanishing operator perturbation \cite{Grace2021}. While this condition is not in general satisfied by the states $\rho_i$ and $\rho_j$ here, the quantum Bhattacharyya bound is always a valid upper bound on the asymptotic error and is often used to approximate the quantum Chernoff bound. Using the quantum Bhattacharyya bound, Eq.~\eqref{eq:GeneralQCE} in the main text can be approximated as
\begin{equation}
	\begin{split}
		\xi_{\textrm{Q},i,j}^{(1)}\approx&\frac{1}{2}\Big[\big(\sqrt{m_{i,x^2}}-\sqrt{m_{j,x^2}}\big)^2\Gamma_{x^2}\\
		&+\big(\sqrt{m_{i,y^2}}-\sqrt{m_{j,y^2}}\big)^2\Gamma_{y^2}\Big]\gamma^2+O(\gamma^3)
	\end{split}
	\label{eq:QBB_Mary}
\end{equation}
This form reveals the dependence of the $M$-ary QCE on differences of square roots of object second moments, constituting a distance measure for quantum-optimal object discrimination.

\begin{figure}[b]
	\centering
	\includegraphics[width=\columnwidth]{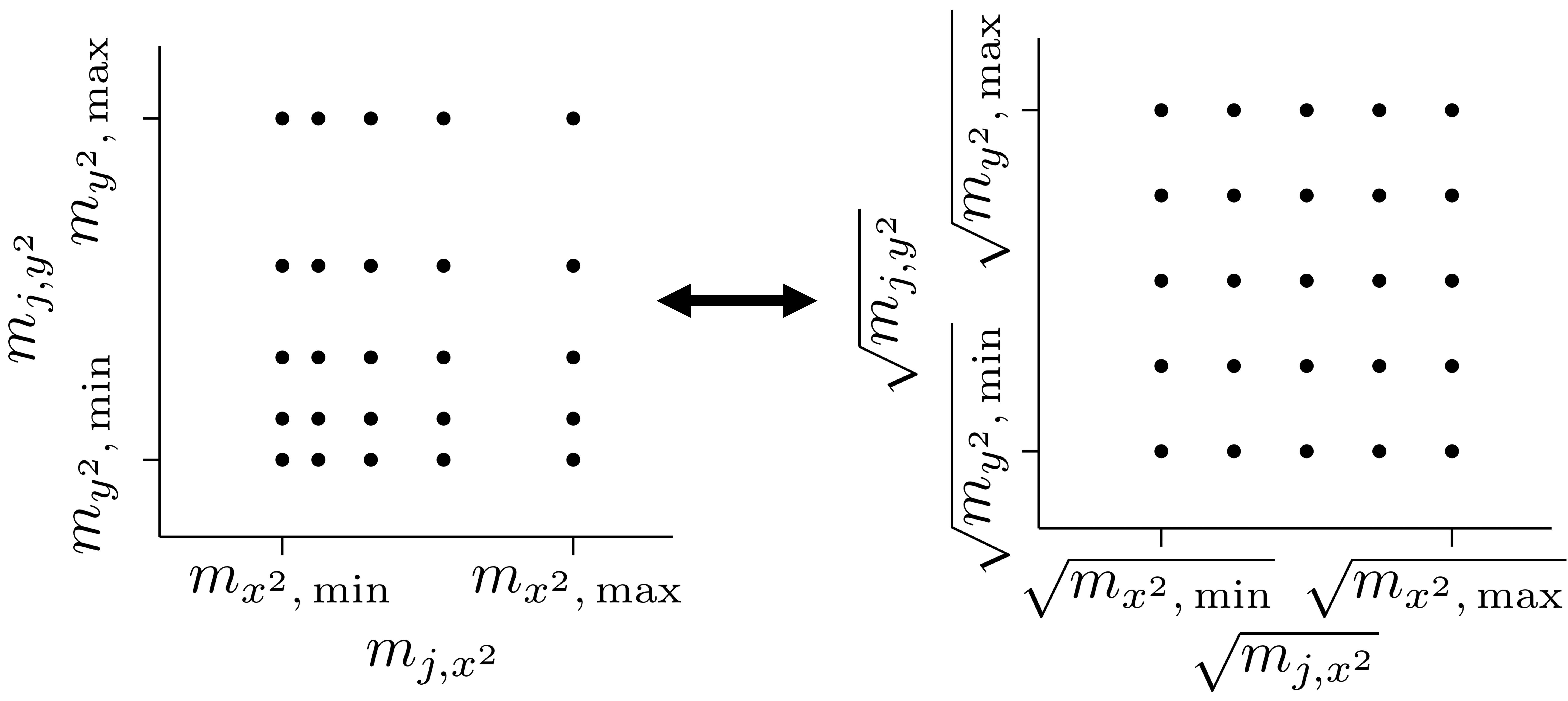}
	\caption{Depiction of an object database with quadratic packing of objects on a rectangular grid. The two representations are equivalent via a coordinate transformation.}
	\label{fig:quadraticPacking}
\end{figure}

First, consider $M$ objects whose 2D second moments are quadratically spaced on a rectangular grid, which is equivalent to requiring that the square roots of the second moments are equidistantly packed in two transverse directions $x$ and $y$ (Fig.~\ref{fig:quadraticPacking}). Formally, this condition can be described by the pair of equations
\begin{equation}
	\begin{split}
		\sqrt{m_{i,x^2}}-\sqrt{m_{j,x^2}}=&(l_{i,x}-l_{j,x})\mu_{\sqrt{x}}\\
		\sqrt{m_{i,y^2}}-\sqrt{m_{j,y^2}}=&(l_{i,y}-l_{j,y})\mu_{\sqrt{y}},
	\end{split} 
	\label{eq:quadraticCondition}
\end{equation}
where $\mu_{\sqrt{x}}=(\sqrt{m_{x^2,\rm max}}-\sqrt{m_{x^2,\rm min}})/(M_x-1)$ and $\mu_{\sqrt{y}}=(\sqrt{m_{y^2,\rm max}}-\sqrt{m_{y^2,\rm min}})/(M_y-1)$ represent the quadratic grid spacing in each direction and where $l_{i,x}\in[1,M_x]$ and $l_{i,y}\in[1,M_y]$ are integer-valued indices for the column and row, respectively, of the $i^{\rm th}$ object in the 2D grid of second moments. It is apparent that the minimal pairwise QCE will occur either when $l_{i,x}-l_{j,x}=1$ and $l_{i,y}=l_{j,y}$ or when $l_{i,x}=l_{j,x}$ and $l_{i,y}-l_{j,y}=1$; the $M$-ary QCE therefore becomes
\begin{equation}
	\xi_{\textrm{Q},M}^{(1)}\approx\frac{1}{2}\min\Big[\mu_{\sqrt{x}}^2\Gamma_{x^2},\mu_{\sqrt{y}}^2\Gamma_{y^2}\Big]\gamma^2+O(\gamma^3).
	\label{eq:QBB_Mary_quadratic2}
\end{equation}
Ensuring that the $x$ direction satisfies the minimum, e.g., by rotating the coordinate system, results in Eq.~\eqref{eq:QCE_M_quadratic} in the main text.

Alternatively, a linearly packed 2D grid of object second moments (Fig.~\ref{fig:regionsHatches}b.) is defined by the condition
\begin{equation}
	\begin{split}
		m_{i,x^2}-m_{j,x^2}=&(l_{i,x}-l_{j,x})\mu_{x}\\
		m_{i,y^2}-m_{j,y^2}=&(l_{i,y}-l_{j,y})\mu_{y}
	\end{split} 
	\label{eq:linearCondition}
\end{equation}
with linear grid spacing $\mu_{x}=(m_{x^2,\rm max}-m_{x^2,\rm min})/(M_x-1)$ and $\mu_{y}=(m_{y^2,\rm max}-m_{y^2,\rm min})/(M_y-1)$.
Under this condition, the pairwise QCEs [Eq.~\eqref{eq:QBB_Mary}] become
\begin{equation}
	\begin{split}
		\xi_{\textrm{Q},i,j}^{(1)}\approx&\frac{1}{2}\Bigg[\bigg(\frac{(l_{i,x}-l_{j,x})\mu_{x}}{\sqrt{m_{i,x^2}}+\sqrt{m_{j,x^2}}}\bigg)^2\Gamma_{x^2}\\
		&+\bigg(\frac{(l_{i,y}-l_{j,y})\mu_{y}}{\sqrt{m_{i,y^2}}+\sqrt{m_{j,y^2}}}\bigg)^2\Gamma_{y^2}\Bigg]\gamma^2+O(\gamma^3).
	\end{split}
	\label{eq:QBB_Mary_Linear}
\end{equation}
By inspection, the minimized pairwise QCE will occur when $l_{i,x}=M_x$, $l_{j,x}=M_x-1$ and $l_{i,y}=l_{j,y}=M_y$ or when $l_{i,x}=l_{j,x}=M_x$, $l_{i,y}=M_y$ and $l_{j,y}=M_y-1$. The resulting approximation to the $M$-ary QCE is
\begin{equation}
	\begin{split}
		\xi_{\textrm{Q},M}^{(1)}\approx&\frac{1}{2}\min\Bigg[\bigg(\frac{\mu_{x}}{\sqrt{m_{x^2,\rm max}}+\sqrt{m_{x^2,\rm max}-\mu_x}}\bigg)^2\Gamma_{x^2}\\
		&+\bigg(\frac{\mu_{y}}{\sqrt{m_{y^2,\rm max}}+\sqrt{m_{y^2,\rm max}-\mu_y}}\bigg)^2\Gamma_{y^2}\Bigg]\gamma^2+O(\gamma^3).
	\end{split}
	\label{eq:QBB_Mary_Linear2}
\end{equation}
Specifying a large object database, i.e., $M_x\gg 1$ and $M_y\gg 1$, implies $\mu_{x}\ll m_{x^2,\rm max}$ and $\mu_{y}\ll m_{y^2,\rm max}$, which can be used to simplify the denominators in Eq.~\eqref{eq:QBB_Mary_Linear2}. An appropriate choice of coordinate axes then yields Eq.~\eqref{eq:QCE_M_linear} in the main text.


For comparison with direct imaging, the exact pairwise CEs $\xi_{\textrm{Direct},i,j}^{(1)}$ from Eq.~\eqref{eq:CEDirectGeneral} can be minimized over all pairs of objects to find $\xi_{\textrm{Direct},M}^{(1)}$. The pairwise direct imaging CEs depend on differences of object second moments in $x$ and $y$, forming a distance measure that is different from that for the quantum limit, as described in the main text. Under a linearly packed 2D object database [Eq.~\eqref{eq:linearCondition}], the pairwise CEs will be minimized either when $l_{i,x}-l_{j,x}=1$ and $l_{i,y}=l_{j,y}$ or when $l_{i,x}=l_{j,x}$ and $l_{i,y}-l_{j,y}=1$, resulting in 
\begin{equation}
	\xi_{\textrm{Direct}}^{(1)}=\frac{1}{32}\min\big[\mu_x^2\Psi_{x^2},\mu_y^2\Psi_{y^2}\big]\gamma^4+O(\gamma^5),
	\label{eq:CE_Mary_Linear}
\end{equation}
where $\Psi_{x^2}=\iint_{-\infty}^{\infty}\psi_{x^2}(\vec{x})^2/\abs{\tilde{\psi}(\vec{x})}^2 d^2\vec{x}$ and $\Psi_{y^2}=\iint_{-\infty}^{\infty}\psi_{y^2}(\vec{x})^2/\abs{\tilde{\psi}(\vec{x})}^2 d^2\vec{x}$. \newpage An appropriate choice of coordinate axes gives Eq.~\eqref{eq:CE_M_linear} in the main text.

Finally, under a quadratically packed 2D object database [Eq.~\eqref{eq:quadraticCondition}], the pairwise CEs for direct imaging become
\begin{widetext}
	\begin{equation}
		\xi_{\textrm{Direct},i,j}^{(1)}=\frac{1}{32}\Big[\big((l_{i,x}-l_{j,x})(\sqrt{m_{i,x^2}}+\sqrt{m_{j,x^2}})\mu_{\sqrt{x}}\big)^2\Psi_{x^2}
		+\big((l_{i,y}-l_{j,y})(\sqrt{m_{i,y^2}}+\sqrt{m_{j,y^2}})\mu_{\sqrt{y}}\big)^2\Psi_{y^2}\Big]\gamma^4
		+O(\gamma^5).
		\label{eq:CE_Mary_quadratic}
	\end{equation}
\end{widetext}
The pairwise CEs will be minimized either when $l_{i,x^2}=1$, $l_{j,x^2}=2$ and $l_{i,y^2}=l_{j,y^2}$ or when $l_{i,x^2}=l_{j,x^2}$, $l_{i,y^2}=1$ and $l_{j,y^2}=2$, resulting in
\begin{equation}
	\begin{split}
		\xi_{\textrm{Direct}}^{(1)}=&\frac{1}{32}\min\Big[(2\sqrt{m_{x^2,\rm min}}+\mu_{\sqrt{x}})^2\mu_{\sqrt{x}}^2\Psi_{x^2}\\
		&+(2\sqrt{m_{y^2,\rm min}}+\mu_{\sqrt{y}})^2\mu_{\sqrt{y}}^2\Psi_{y^2}\Big]\gamma^4+O(\gamma^5).
	\end{split}
	\label{eq:CE_Mary_quadratic2}
\end{equation}
Specifying a large object database implies $\mu_{\sqrt{x}}\ll 2\sqrt{m_{x^2,\rm min}}$ and $\mu_{\sqrt{y}}\ll 2\sqrt{m_{y^2,\rm min}}$, simplifying Eq.~\eqref{eq:CE_Mary_quadratic2}. Choosing the coordinate axes gives Eq.~\eqref{eq:CE_M_quadratic} in the main text.

For completeness, we note that the rectangular grid configurations shown in Fig.~\ref{fig:regionsHatches}a. and Fig.~\ref{fig:regionsHatches}b. are not the optimal way to pack $M$ points in a rectangular region. For example, assuming an object database with $x$-$y$ symmetry, the maximized minimum distance between any two points within a square region is known to have the asymptotic behavior $D_M\sim\sqrt{2/(\sqrt{3} M)}$ as $M\to\infty$ \cite{Croft1991}, whereas the square grid configuration yields $D_M\sim1/\sqrt{M}$. Assuming a circularly symmetric aperture, employing the optimal packing configuration would therefore increase the results in Eqs.~\eqref{eq:QCE_M_quadratic}-\eqref{eq:CE_M_linear} in the main text by a factor of $2/\sqrt{3}\approx1.155$ in the large-database regime. We conclude that considering the optimal configuration of objects does not significantly change our results.
\end{document}